\documentclass[journal]{IEEEtranTCOM}
\usepackage{amsmath}\usepackage{amssymb}\usepackage{graphicx}
\usepackage{dblfloatfix}
\usepackage[caption=false,font=footnotesize]{subfig}

\makeatletter

\usepackage{epsfig}\usepackage{amsfonts}\usepackage{array}\usepackage{dcolumn}\usepackage{psfrag}\usepackage{amsfonts}\usepackage{accents}\usepackage{booktabs}
\usepackage{cite}\usepackage[all]{xy}
\usepackage{dsfont}\usepackage{url}\usepackage{xcolor}\usepackage[nolist]{acronym}
\usepackage[footnote,draft,silent,nomargin]{fixme}
\usepackage{algorithm}
\usepackage[noend]{algpseudocode}
\usepackage{amsthm}
\usepackage{mathtools}
\usepackage{bbm}
\usepackage{xfrac}
\usepackage{booktabs}
\let\labelindent\relax\usepackage{enumitem}
\usepackage{bm}

\graphicspath{{./figures/}}
\acrodef{SINR}{Signal to Interference and Noise Ratio}
\acrodef{CDF}{Cumulative Distribution Function}
\acrodef{PDF}{Probability Density Function}
\acrodef{MTC}{Machine-type Communications}
\acrodef{M2M}{Machine-to-Machine}
\acrodef{CTMC}{Continuous-Time Markov Chain}
\acrodef{RV}{Random Variable}
\acrodef{MTTR}{Mean Time to Restoration}
\acrodef{TTI}{Transmission Time Interval}
\acrodef{SDN}{Software Defined Networking}

\acrodef{ARP}{Access Reservation Protocol}
\acrodef{UE}{User Entitity}
\acrodef{RAO}{Random Access Opportunity}
\acrodef{MTC}{Machine Type Communication}
\acrodef{TTI}{Transmission Time Interval}
\acrodef{LTE}{Long Term Evolution}

\definecolor{darkpink}{rgb}{0.91, 0.33, 0.5}

\definecolor{orange}{RGB}{255, 165, 0}

\newcommand{\bfH}{{\mathbf{H}}}
\newcommand{\bfw}{{\mathbf{w}}}
\newcommand{\bfs}{{\mathbf{s}}}
\newcommand{\bfu}{{\mathbf{u}}}
\newcommand{\bfv}{{\mathbf{v}}}

\newcommand{\bfU}{{\mathbf{U}}}
\newcommand{\bfV}{{\mathbf{V}}}
\newcommand{\bfR}{{\mathbf{R}}}
\newcommand{\bfF}{{\mathbf{F}}}

\makeatother

\begin{document}

%

\title{Wireless Access in Ultra-Reliable Low-Latency Communication (URLLC)}

\author{
	\emph{Invited Paper} \\ \vspace{18pt}
  	\IEEEauthorblockN{%
  	Petar~Popovski, 
  	\v Cedomir~Stefanovi\' c, 
  	Jimmy~J.~Nielsen, 
  	Elisabeth~de~Carvalho, 
  	Marko~Angjelichinoski,
  	Kasper~F.~Trillingsgaard, and
	Alexandru-Sabin Bana \\
\IEEEauthorblockA{Dept.~of Electronic Systems, Aalborg University, 9220 Aalborg, Denmark\\
Emails: %
\{petarp,cs,jjn,edc,maa,kft,asb\}@es.aau.dk %
}%
}%
}%
\maketitle
%
%
\begin{abstract}
The future connectivity landscape and, notably, the 5G wireless systems will feature Ultra-Reliable Low Latency Communication (URLLC). The coupling of high reliability and low latency requirements in URLLC use cases makes the wireless access design very challenging, in terms of both the protocol design and of the associated transmission techniques. This paper aims to provide a broad perspective on the fundamental tradeoffs in URLLC as well as the principles used in building access protocols. Two specific technologies are considered in the context of URLLC: massive MIMO and multi-connectivity, also termed interface diversity. The paper  also touches upon the important question of the proper statistical methodology for designing and assessing extremely high reliability levels.  
\end{abstract}
\begin{IEEEkeywords}
Ultra-reliable communication, URLLC, IoT, 5G, access protocols, masssive MIMO, multi-connectivity.
\end{IEEEkeywords}

\section{Introduction}


During the past three decades wireless connectivity has become a commodity, assumed to be practically always present and visible only when absent. This has naturally increased the confidence in wireless-enabled applications and services, leading to the idea of using wireless at a large scale to support mission-critical communication links. This trend has been termed \emph{ultra-reliable communication (URC)}~\cite{popovski2014ultra}, where the level of connectivity guarantees, e.g. $ > 99.999 \, \%$ of the time, matches the cable-based communication systems. 

Ultra-reliability has inevitably become a part of the emerging 5G wireless systems. Indeed, 5G aims to cover three generic connectivity types: enhanced Mobile Broadband (eMBB), massive Machine-Type Communication (mMTC) and 
Ultra-Reliable Low-Latency Communication (URLLC). As it can be seen from the name, ultra-reliability is entangled with the requirement for low latency in the context of 5G systems. This makes URLLC very challenging, but also rather restrictive. In the earlier days of ultra-reliable wireless~\cite{popovski2014ultra}, there was a proposal to consider two types of ultra-reliable connectivity: \emph{(i)} URC over a long term, in which the required latency is $> 10$~ms; \emph{(ii)} URC in a short term, with latency of $\leq 10$~ms. URC over a long term is interesting for use cases in which one needs \emph{resilient} wireless connections, such as in disaster scenarios or remote interactions with a larger latency budget, e.g. changing a route of a drone. URC over short term contains URLLC\footnote{URLLC is often associated with latencies of around 1~ms, such that $10$~ms in this context is too long.} and is meant for applications with very stringent latency requirements, such as communication among machines and robots in Industry 4.0 use cases. However, while URLLC has been established as a concept in the community, URC over long term has been only scarcely present. We will therefore keep the focus in the paper on URLLC, noting that the insights about ultra-reliable connections and the communication-theoretic principles discussed here can be applied to URC defined over both short and long term.

In this paper we will treat a set of fundamental problems in wireless access for URLLC. The objective is to provide the reader with a framework that can be used to analyze and design ultra-reliable wireless systems. Our previous article~\cite{PNSCSTBKKS2018} can be seen as a predecessor of this work, where we have outlined the principles and the building blocks for wireless access in URLLC. This paper is intended to provide an in-depth treatment of some of the aspects and techniques associated with URLLC. We provide a detailed discussion on the communication-theoretic principles that are underpinning the design of URLLC. Compared to \cite{PNSCSTBKKS2018}, here we have a detailed discussion on medium access control (MAC) protocols, use of large number of antennas in massive MIMO for providing high reliability, as well as the concept of interface diversity and multi-connectivity. We are also addressing a fundamental question, largely ignored in the literature so far: what are the statistical requirements to measure and verify ultra-reliability. It should be noted that this paper does not include all the details relevant for the discussion on the transmission of short packets; this has been discussed to a sufficient level in~\cite{PNSCSTBKKS2018} and~\cite{durisi2015towards}. 

The paper is organized as follows. The next section provides an overview of the URLLC use cases that create the context for developing wireless access protocols, as well as the requirements associated with them. Section~\ref{sec:CT} elaborates on the communication-theoretic principles of URLLC, providing a perspective on the relationship between latency, packet size, bandwidth, and finite-blocklength treatment. This is followed by Section~\ref{sec:AP} on access networking, where a special emphasis is put on the problem of frame synchronization, a procedure that needs to have very high reliability in order to support packet decoding in URLLC scenarios. Section~\ref{sec:MassiveMIMO} sheds light on Massive MIMO, a technology that relies on extreme spatial diversity, which makes it a natural candidate for supporting ultra-reliable transmissions. 
Since the future URLLC devices are likely to have multiple communication interfaces, Section~\ref{sec:IFD} is dedicated to ultra-reliability achieved through multi-connectivity, i.e. interface diversity. Section~\ref{sec:LTK} treats the fundamental questions related to the statistical aspects of ultra-reliability. Some references to the related work are scattered through the text, while a comprehensive overview of the state-of-the-art in the URLLC literature is provided in Section~\ref{sec:RelatedWork}. The last section concludes the paper and provides a perspective on some open issues. 

\section{URLLC Use Cases and Requirements}
\label{sec:UCR}

URLLC brings a significant novelty to 5G as a system. Along with mMTC, it makes 5G qualitatively different from the previous mobile wireless generations. Ultra-reliable communication is potentially an enabler of a vast set of applications, some yet unknown. To put this in perspective, wireless connectivity and embedded processing have significantly transformed many products by expanding functionality and transcending the traditional product boundaries~\cite{porter2014smart}. For example, a  product stays connected to its manufacturer through its lifetime for maintenance and update.
Ultra-reliable wireless brings this transformation to the next level, as the availability of wireless connectivity practically all the time is an important assumption that a system designer should account for when designing a system. For example, ultra-reliable wireless connectivity between two parts of a system removes the need for their physical attachment. 

In general, the applications and the use cases of URLLC can be divided into two groups: \emph{(i)} cable replacement and its extensions and \emph{(ii)} native URLLC applications. The ones related to cable replacement are transforming some of the current applications that rely on cabled connections, but also add a new quality due to the flexibility of wireless. An example of this are the digital systems in Industry 4.0, where wireless will replace cabled connections, but also give rise to new types of interactions, e.g. among cooperative robots. On the other hand, a \emph{native} URLLC application is the one that has no precedent in wired communication; an example is vehicle-to-vehicle (V2V) communication.   

A comprehensive treatment of the URLLC use cases is carried out in 3GPP standardization. At the moment of writing of this article, 3GPP is about to start standardization work on Release 16, which should address both the reliability and latency in future mobile cellular networks, thus setting the stage for URLLC services.
The general vision of URLLC requirements by 3GPP is presented in \cite{TR38.913}:
\begin{itemize}
    \item A reliability requirement of $1-10^5$ (i.e. 99.999 \%) with a user-plane radio latency\footnote{Radio latency is measured from the moment of the reception of a packet by layer-2 radio protocol at the transmitting end to the moment of the delivery of the packet to the layer-3 protocol at the receiving end.} of 1 ms for a single transmission of 32-byte long packet.
    \item A user-plane average latency of 0.5 ms for both uplink and downlink, without an associated reliability value.
\end{itemize}
However, these figures are by far insufficient to describe the variety of use cases and the associated requirements of the verticals that 5G is envisioned to support, as discussed next. Furthermore, as discussed in Section~\ref{sec:LTK}, these specifications are also insufficient from a statistical viewpoint.

The automotive 5G URLLC cases represent an important segment of the ongoing 3GPP standardization and can be divided into \emph{assisted}, \emph{co-operative} and \emph{tele-operated driving}~\cite{D2.1,TR22.886}.
Their user plane reliability requirement is $1-10^5$ with the associated maximum end-to-end (E2E) latency requirement of 5~ms for assisted, 10~ms for co-operative, and 20~ms for tele-operated driving, both in the uplink and downlink.
Note that, as a rule of thumb, radio latency can be estimated as 1/10 of E2E latency~\cite{D2.1}.

Another important set of URLLC use cases is related to monitoring and control of industrial processes, belonging to the emerging paradigm of Industry~4.0.
The most important examples are \emph{motion control}, \emph{factory automation} and \emph{process automation}~\cite{D2.1,TS22.261}.
Motion control pertains to real-time control of machines with moving parts, and is characterized by user-plane reliability of $1-10^{-5}$ with E2E latency of 1~ms (i.e. the user-plane radio latency of 0.1~ms).
Moreover, this use case is about isochronous transmission of sensory and actuation information in the upink and downlink, respectively, requiring user-plane E2E jitter of 1~{\textmu}s.
Factory automation (also referred to as discrete automation or discrete manufacturing), according to 3GPP~\cite{TS22.261}, requires user-plane reliability of $1 - 10^{-4}$ with user-plane E2E latency of 10~ms and jitter of 100~{\textmu}s.
However, in some other sources, this use case is characterized with an extreme reliability requirement of $1-10^{-9}$ or more~\cite{ericsson, siemens, CASHBLV2018, HWWTAHAA2016}, with a more demanding user-plane latency of 1~ms (for local monitoring and control setups) and 5~ms for (remote setups) and jitter of 1~{\textmu}s~\cite{siemens}.
Process automation, which is related to production of goods in bulk quantities, requires user-plane reliability of $1-10^{-6}$ and E2E latency of 50~ms and jitter of 20~ms, according to 3GPP~\cite{TS22.261}.
Again, industrial sources aim at more stringent values that match the ones for the factory automation~\cite{ericsson, siemens}.

We also mention the category of URLLC use cases that belong to the \emph{tactile Internet}; their common feature is the existence of haptic feedback which puts the most stringent requirements in terms of reliability and latency.
As an example, the haptic feedback in tele-surgery may require reliability of $1-10^{-9}$ and round-trip time as low as 1~ms~\cite{CASHBLV2018}.

The novelty of 5G is that reliability and latency are also explicitly involved in mMTC use cases, e.g. in monitoring of non-time critical process and logistics in the contexts of smart cities and factories~\cite{D2.1}, where user-plane reliability is set to 95~\% with a maximum radio latency of 0.5~ms.
Moreover, enhanced Mobile BroadBand (eMBB) service category also features general requirements of user-plane radio latency of 4~ms, both in the uplink and downlink~\cite{TR38.913}.
These latency figures are lower than what 4G is able to provide, where the target user-plane radio latency is 10~ms~\cite{M.2134}.
We also note that reliability, as defined in 5G standardization, does not exist as a requirement in 4G. 
In summary, low latency and high reliability seem to be intrinsic to 5G, no matter the actual use case and service category.

Finally, we note that in this section we have focused only on the latency and reliability as the key performance parameters.
More information about other performance parameters, such as availability, experienced data rates, payload sizes, as well as about deployment setups, security and other features, can be found in the references mentioned in the section.

\section{Communication-Theoretic Principles of URLLC}
\label{sec:CT}

The objective of this section is to introduce communication-theoretic considerations on the modeling and the fundamental tradeoffs in URLLC.


\subsection{Communication-Theoretic Model}

We will build our discussion of design principles and analysis based on the following baseband model of a received signal $y$ 
\begin{equation} \label{eq:BasicCommModel}
y=h \, \alpha \, x + z + w
\end{equation}
as well as its generalizations. Here  $h$ is the channel coefficient, which in the general MIMO case is a matrix of channel coefficients; $\alpha$ is the activity indicator; $x$ is the transmitted signal; $z$ is the noise; and $w$ is the interference. The activity indicator value is $\alpha=1$ if there is an actual transmission $x$ and is $\alpha=0$ otherwise. All variables $h$, $\alpha$, $x$, $z$, $w$ are random and contain uncertainty; however, the receiver wishes to learn only $\alpha$ and, if $\alpha=1$, decode $x$.
The knowledge about the other three variables $h,w,z$ can be partial, statistical, or even non-existing. Let us take an initial look into the nature of these random variables; we will treat $h,\alpha$ and $x$ in details throughout the paper.  

The most common random disturbance in communication systems is the noise $z$. The statistics of the noise is usually known and in the most common case is Gaussian, with a known noise power. Some of the most fundamental results in information theory, both in asymptotic case and in the case of packets with finite blocklength, are related to the Gaussian channel with known noise variance.

The situation is substantially different when the interference term $w$ is considered.
The knowledge about $w$ depends on the part of the radio spectrum in which the bandwidth $B$ is allocated. In a spectrum that has a certain type of license, the license-owner pays in order to acquire the right to manage the interference in that spectrum. This does not mean that the interference is non-existent, but is turned into a \emph{known unknown} and the spectrum owner can control or at least influence the interference and its statistics.  

On the other hand, if the spectrum is unlicensed, then the statistics of $w$ is largely unknown.
Indeed, the open access to the unlicensed spectrum puts constraints on the way a given transmitter may operate, but does not limit the number of independently owned systems that can run in close proximity of each other\footnote{In other words, one can buy and turn on an arbitrary number of WiFi access points in a small space, e.g. room and set them up to transmit at different channels, thereby occupying the whole unlicensed spectrum.}. The interference in unlicensed, but also sometimes in licensed bands, can be regarded as the most significant ``unknown unknown'' in the system model and one should used risk-based methods~\cite{bennis2018ultra} to assess its impact for URLLC communication. This is elaborated in Section~\ref{sec:LTK}.

The knowledge of the channel $h$ or at least its statistics is critical in URLLC systems. Even if we consider a non-coherent communication, where the receiver does not need to know or to learn $h$, the precise knowledge of the statistics of $h$ is crucial to be able to guarantee a certain reliability of communication. 

Finally, finding out $x$ is the central task of each receiver and we will treat it throughout the whole paper. The level of knowledge about the activity of the transmitter $\alpha$ depends on the communication scenario. In a downlink transmission, the BS is the only transmitting candidate (except in a discovery process) and the receiving device expects to receive the signal, such that for this case we can take $\alpha=1$. However, for uplink transmission, in general, the BS does not a priori know whether the user is active, which translates into uncertainty about $\alpha$.
Finding out the values $\alpha$ for the devices connected to the same BS is the access protocol problem, treated in Section~\ref{sec:AP}, and it contributes substantially to the ultra-reliable performance.  

In the rest of the paper, we will treat in details  various techniques and aspects of URLLC in the light of the model given by~\eqref{eq:BasicCommModel}.

\subsection{Relating Latency and Reliability}

Latency can be defined in different ways and at different layers of the communication protocols. The simplest definition of a latency, treated in this paper, is the delay that a data packet experiences from the ingress of a given protocol layer at the transmitter to the egress of the same layer at the receiver. In applications related to, e.g. remote controls of robots or drones, one is interested in a two-way or round-trip delay. 

\begin{figure}
    \centering
    \includegraphics[width=8.3cm]{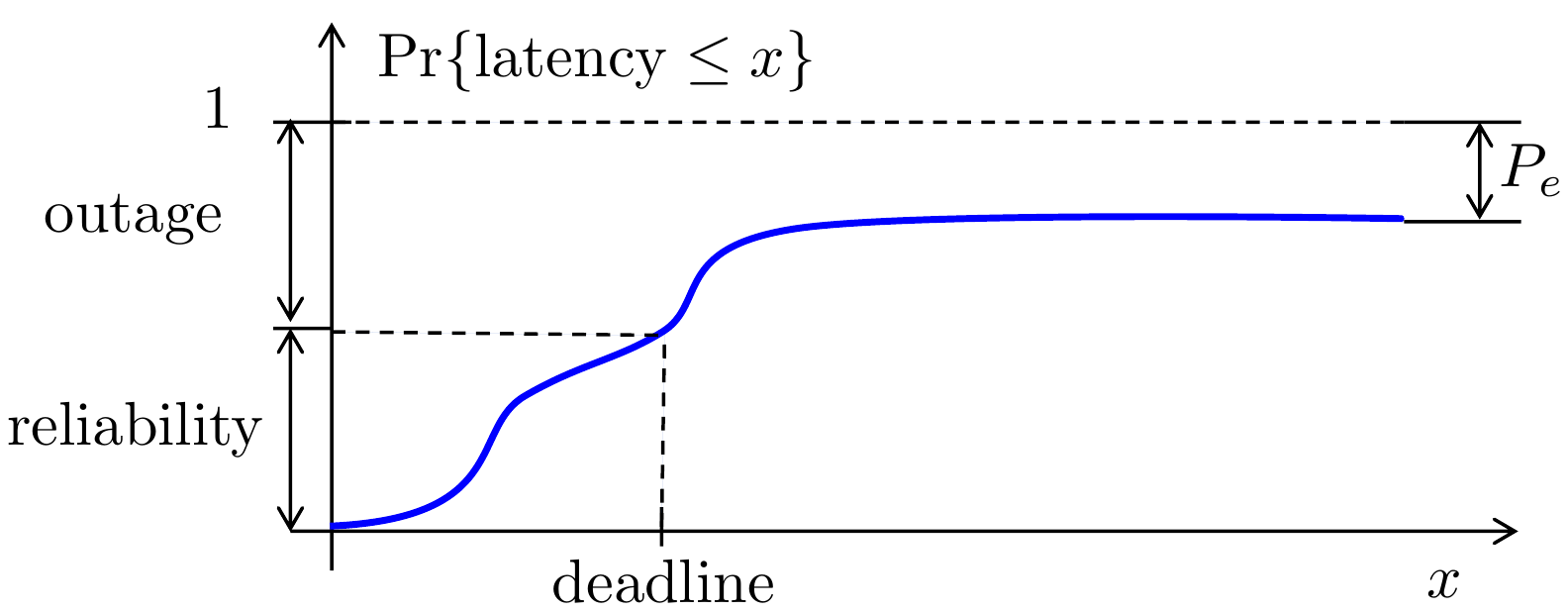}
    \caption{Relation between outage, reliability, latency, and deadline.}
    \label{fig:LatencyReliability}
\end{figure}

Under the constraints of a URLLC service, the definition of reliability should be coupled to the latency requirement. In fact, one can say that, when the latency requirement is absent (theoretically infinite), then transmitting at a rate that is lower than channel capacity offers perfect reliability. From the perspective of an application, with a predefined latency constraint, we can define the \emph{reliability} of a communication setup as the probability that the latency does not exceed this deadline, and \emph{outage} as the probability that it does. Fig.~\ref{fig:LatencyReliability} shows the generic requirement in terms of latency and reliability, applicable not only to point-to-point link, but also arbitrary communication setup. The exact numbers on the deadline and the reliability are dependent on the application. 
We note that the latency cumulative distribution function (CDF) asymptote is equal to $1-P_e$, where $P_e$ is the probability of residual packet loss or packet error. This residual packet loss reflects the fact that some packets will never be delivered  due to, for example, limits on the number of retransmissions in link-layer protocols, buffer overflows, synchronization failures, etc. 

\subsection{The Fundamental Tradeoffs and Packet Structure}

As URLLC is often associated with transmission of controls and commands over wireless link in a distributed system, one of the basic assumptions about URLLC is that they involve small payloads. This naturally creates the association with the transmission of short packets~\cite{durisi2015towards} and the use of finite-blocklength information theory. It is instructive to look at the basic choices and tradeoffs that decide the packet length in an URLLC setting. 

Instead of talking about the payload size and latency, we consider a set of five variables: bandwidth, rate, reliability, energy, and latency. Let us at first fix the latency to $T$. Given the payload size of $D$ bits and the maximal latency $T$, we can determine the minimal transmission rate $R_{\mathrm{bps}}$ in [bps]. Note that in the standard information-theoretic models, the data rate is expressed in terms of bits per channel uses [bpcu], here denoted simply by $R$. By selecting the bandwidth $B$, the number of channel uses available for transmission is $2BT$, such that the different types of data rates are related as follows:
\begin{equation}
R=\frac{D}{2BT}=\frac{R_{\mathrm{bps}}}{2B} \quad \mathrm{[bpcu]}
\end{equation}
The next variable is the energy. This may refer to the total energy consumed during the transmission by the transmitter and the receiver. It can even be understood in a more general way, encompassing anything that adds diversity and hardens the received signal, such as the use of multiple antennas (however, without adding spatial channels for additional data multiplexing). Let us fix the energy used for transmission within the time $T$. The SINR at the receiver is determined by this energy, along with the channel realization and the interference, which are variables that cannot be chosen. With all these variables fixed, one can determine the achievable reliability of the transmission, denoted by $1-\epsilon$. In an analogous way, of other four variables are fixed, for example rate, reliability, energy and latency, then one can find what is the required bandwidth $B$. 

These, rather basic, considerations, are very important to get clarity in describing the models for URLLC. This is illustrated by the following two URLLC aspects:

\noindent \emph{(1)} Given the latency $T$, the size of the packet blocklength in terms of available channel uses, equal to $N=2BT$, can be regulated by selecting the bandwidth $B$. If the bandwidth available for transmission is very large, then the blocklength becomes very large as well. In other words, large bandwidth can move the transmission regime towards asymptotically large packet lengths; however, the data rate becomes very low and so does the spectral efficiency. 

\noindent \emph{(2)} Both the sender and the receiver use energy during the communication. Assume there is a single sender, Alice, and two possible receivers, Bob and Carol. If Alice sends to Bob, but not to Carol, then the activity indicator for Bob, denoted by $\alpha_B$, see (\ref{eq:BasicCommModel}), is given by $\alpha_B=1$. For Carol it should be $\alpha_C=0$. If this is not known in advance, e.g. through pre-scheduling, then Bob and Carol should learn it from Alice's transmission, which requires spending some energy on detection and decoding and carrying out a hypothesis testing about the activity factors $\alpha_B$ and $\alpha_C$. Alternatively, they can both set always $\alpha_B=\alpha_C=1$ and receive anything that comes from Alice. In this case, only after decoding the packet, Bob and Carol figure out who is the intended recipient of the packet. This can improve the reliability, since $\alpha$ does not need to be decided separately, but it also increases the receiver energy consumption, as a receiver decodes packets that are not necessarily intended for him/her. 

\subsection{URLLC Packet Structure}
\label{eq:URLLCpacket}

As already mentioned, a general requirement for URLLC in 3GPP is reliability of $1 - 10^{-5}$ (i.e. probability of error $\epsilon=10^{-5}$) with latency of $T=1$~ms and for a transmission of a packet of size $D=32$ bytes.
In addition to the data, the packet should also contain signaling information/metadata. 

\begin{figure}[t]
    \centering
    \includegraphics[width=\linewidth]{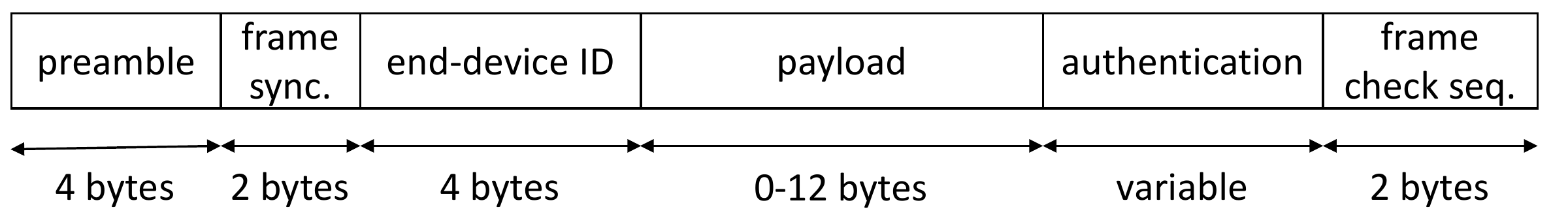}
    \caption{Example of a packet format used in a low-throughput IoT system. The structure is largerly inherited from the common packet structure used in broadband systems.}
    \label{fig:LTNpacket}
\end{figure}

Although not directly related to URLLC, a good example of a short packet format is depicted on Fig.~\ref{fig:LTNpacket}, taken from low throughput networks~\cite{ETSI2014LTN}. It clearly illustrates that a significant portion of the packet is spent on metadata as well as resources for performing auxiliary operations, such as synchronization and packet detection.
As discussed in \cite{PNSCSTBKKS2018}, when the reliability requirements are as high as in URLLC, 
one can no longer assume that the transmission of the metadata and the auxiliary procedures are perfectly reliable. Indeed, the probability of success for a given packet $\pi$ with a structure as the one on Fig.~\ref{fig:LTNpacket} is given by:
\begin{equation} \label{eq:SeparateEncoding}
P_S(\pi)=P_S(A) P_{S}(M) P_{S}(D)
\end{equation}
where $P_S(A), P_{S}(M)$, and $P_{S}(D)$ denote the success probability of the auxiliary procedures, metadata and data, respectively. This illustrates the point that the packet design that is based on separation of the resource for auxiliary procedures, metadata, and data leads to product of the success probability of the different elements, thus deteriorating the overall reliability. 
\begin{figure}[t]
\centering
\includegraphics[width=\linewidth]{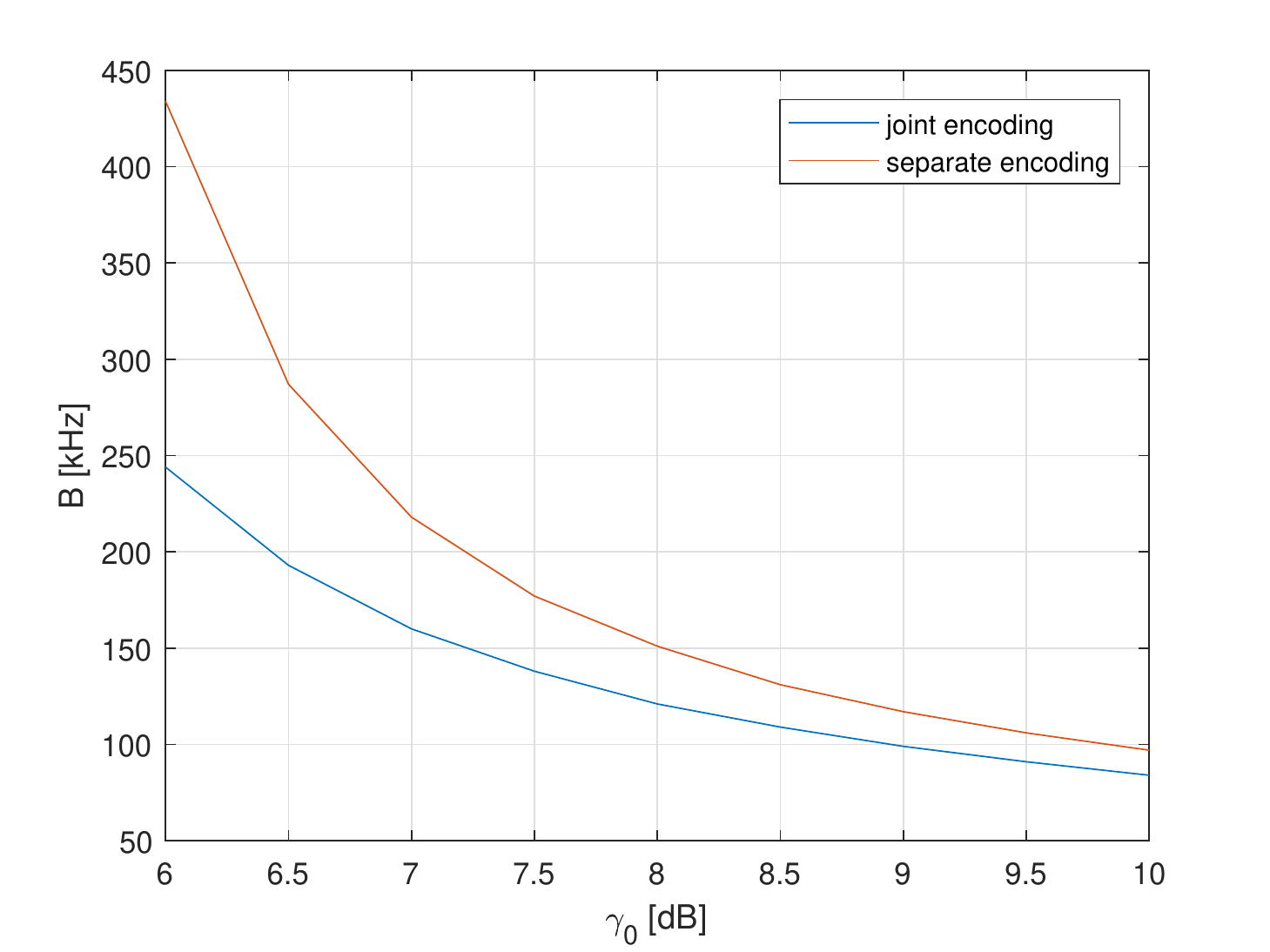}
\caption{Minimal required bandwidth to send $16$ bytes of data and $16$ bytes of metadata for the case of separate and joint encoding, respectively. The latency is set to $T=1$~ms and the reliability to $1-10^{-5}$. Here $\gamma_0$ is the reference SNR for a bandwidth of $B_0=100$ kHz; the SNR for a bandwidth $B$ is given by $\gamma_B=\gamma_0 \dfrac{B_0}{B}$.}
\label{fig:BW_SNR_URLLC}
\end{figure}

Let us illustrate the impact of (\ref{eq:SeparateEncoding}) for short packets. For simplicity, let us assume that $P_S(A)=1$, while the packet has $D=16$ bytes of data and $M=16$ bytes of metadata. The latency is set to $T=1$ ms, while the bandwidth $B$ should be determined such as to achieve the desired reliability of $1-10^{-5}$. For the sake of argument, we assume that $B$ is lower than the coherence bandwidth of the system, such that a single coefficient describes the channel and the SNR achieved at the receiver. The channel coefficient is assumed to be known, such that the finite blocklength bounds for the complex AWGN channel hold. Hence, the error probability of receiving $b$ bits of data within $N=2BT$ channel uses when the SNR is given by $\gamma$ is well approximated by \cite{finite_blocklength_polyanskiy} \cite{durisi2015towards} 
\cite{bana_short_pkt_tradeoff}:
\begin{align}
    \epsilon(N,\gamma,b) = \text{Q} \mathopen{} \left(\dfrac{N C(\gamma) - b + \frac{1}{2} \log_2 N }{\sqrt{N V(\gamma)} } \right)
    \label{eq:error_finite_blocklength}
\end{align}
where  ${C(\gamma)= \dfrac{1}{2} \log_2(1+\gamma )}$ and ${V\mathopen{}(\gamma)=\dfrac{\gamma(\gamma+2)}{2(\gamma+1)^2}} \log_2^2 e$ denote the channel capacity and dispersion, respectively. Note that the SNR $\gamma$ also depends on the bandwidth $B$ and thereby on $N$, the number of available channel uses. Indeed, let us fix a reference SNR to $\gamma_0$ for some bandwidth $B_0$. Then, for equal transmit power (i.e. useful received power), the SNR when the bandwidth is $B$ is given by 
\begin{equation}\label{eq:gammavsgamma0}
\gamma=\gamma_0 \frac{B_0}{B}= \gamma_0 \frac{2B_0T}{N}.
\end{equation}

We consider two cases:
\begin{enumerate}
\item Data and metadata are encoded jointly, the probability of correct packet reception is $(1-\epsilon(N,\gamma,D+M))$;
\item Data and metadata are encoded separately and the probability of correct packet reception is $[1-\epsilon(N/2,\gamma,M)][1-\epsilon(N/2,\gamma,D)]$.
\end{enumerate}
Fig.~\ref{fig:BW_SNR_URLLC} shows the required bandwidth $B$ as a function of the reference SNR $\gamma_0$ in order to achieve the required reliability with the prescribed latency. The reference SNR $\gamma_0$ in (\ref{eq:gammavsgamma0}) is fixed to $B_0=100$~kHz. Clearly, when the data and the metadata are jointly encoded, the required number $N$ of channel uses is lower compared to the case when they are encoded separately. This results in a lower required bandwidth, as the figure shows.\footnote{The bandwidth values shown on Fig.~\ref{fig:BW_SNR_URLLC} are too large for a commonly seen values of coherence bandwidth and here they serve for illustrative purpose only.} 

The trick with increasing the blocklength in order to attain a more efficient transmission, given the reliability and the latency constraints, can also be used in a scenario in which a Base Station (BS) broadcasts to multiple terminals, as in \cite{trillingsgaard2017downlink}. Namely, given the total number of channel uses for broadcast, the BS can concatenate all packets intended for different users and use the resulting large packet as a input to the encoder. Intuitively, this offers the highest reliability, but the price is that each node needs to spend energy to decode data that it does not need, as it decodes the whole packet before seeing if there is any data intended for that node and, if yes, extract it. The tradeoff between the reliability and energy expenditure is analyzed in \cite{trillingsgaard2017downlink}. This technique has been termed \emph{concatenate-and-code} in~\cite{tuninetti2018scheduling}, which has extended the work towards cross-layer scheduling.  

Nevertheless, the same trick of increasing the blocklength by aggregating data cannot be used when different data chunks are transmitted by different nodes. In other words, if the BS has a data chunk for Alice and Alice has a data chunk from the BS, it is not possible to aggregate both data chunks in order to counter the effect of finite blocklength. 

In the most ``honest'' case for URLLC transmission, the receiver does not have the channel state information. If a coherent transmission is about to take place over bandwidth that is larger than the coherence bandwidth, then the receiver should use pilots to estimate the channel and these pilots consume significant resources. Alternatively, the transmission can be carried out in a non-coherent way. Rigorous studies that take a holistic view on channel estimation and packet transmission are presented in \cite{ostman2017short,ferrante2018pilot}.

In summary, increased bandwidth is one of the most straightforward ways to bring diversity to the URLLC transmissions. There are two mechanisms that contribute to it, depending on the relationship between the transmission bandwidth $B$ and the coherence bandwidth of the system $B_c$. 
\begin{itemize}
\item $B \leq B_c$: In this case, the SNR of all channel uses/symbols is identical. If the data $D$ to be transmitted is fixed, then reliability increases due to the lower rate per channel use (symbol). On the other hand, the transmission blocklength becomes longer, which improves the transmission efficiency according to the results of the finite blocklength information theory. 
\item $B > B_c$: In this case the channel uses that are frequency-separated for more than $B_c$ have different SNR statistics, which brings a new degree of diversity, in addition to the one brought by the increased number of channel uses. 
\end{itemize}

Regarding the resource allocation in practical 5G systems, the tradeoffs arising in relation to the definition of time-frequency resources is captured in the flexible numerology used to design the 5G frames~\cite{berardinelli2015design}.

\section{Access Networking}
\label{sec:AP}

The set of physical, MAC and link-layer protocols are referred to as \emph{access networking}, which has the following tasks: \emph{(i)} resolving the uncertainty in the user activity and inferring the value of $\alpha$ from (\ref{eq:BasicCommModel}); \emph{(ii)} performing the auxiliary procedures, notably synchronization; \emph{(iii)} decoding metadata and data; and \emph{(iv)} interacting with the higher-layer protocols, where latency is ultimately measured and assessed.
Its operation has to be designed according to the target reliability-latency requirements and to the traffic patterns of the supported services.
In this section we consider several generic access networking options for URLLC services.

\subsection{Access Networking for URLLC Services with Deterministic Traffic Arrivals}

URLLC services with deterministic traffic arrivals pertain to closed-loop control applications that involve deterministic sensing-actuation cycles with rather short periods and extreme latency-reliability requirements.
The examples of such services, which in essence demand isochronous communications, are motion control and factory automation.

For deterministic traffic arrivals, the value of $\alpha$ in \eqref{eq:BasicCommModel} is known a priori.
A sensible access networking approach in this case is to employ periodic, pre-configured reservation of resources for data transmissions in both uplink and downlink, providing a deterministic timing of the traffic exchanges.
This static allocation of the resources could be done offline, or using signaling exchanges that take place before the execution of the service starts.

The operation of access networking with static allocation of resources relies critically on precise time synchronization\footnote{Time synchronization relates to the distribution of absolute time references~\cite{ITU-T}.} among all involved network elements, which could be achieved using an external synchronization network, such as GPS.
If such solution is not viable, due to, e.g. cost or indoor/obscured device location, one could employ synchronization methods that are reliant on dedicated signaling exchanges among the network elements.
Nevertheless, achieving and maintaining such high level of synchronism in this way is a challenging task, particularly if the jitter requirements of the service are stringent, which typically is the case, see Section~\ref{sec:UCR}.

The error probability of a packet exchange between a device and the Base Station (BS) in case of static allocation is given by
\begin{equation}
\label{eq:prealloc}
\epsilon_{\textrm{det}} = 1 - ( 1 - \epsilon_\text{sync} ) ( 1 - \epsilon_D) ( 1 - \epsilon_A)
\end{equation}
where $\epsilon_\text{sync}$ is the synchronization error, and $\epsilon_D$ and $\epsilon_A$ are the probabilities that the data and the acknowledgement are not successfully decoded, respectively, which depend only on the choice of transmission parameters.
Obviously, the used synchronization and the transmission methods have to ensure that
$\epsilon < \epsilon_\text{target}$, where $\epsilon_\text{target}$ is the target error-probability of an acknowledged packet exchange.

\subsection{Access Networking for URLLC Services with Stochastic Traffic Arrivals}
\label{sec:deterministic}

In case of services with stochastic traffic arrivals, the value of $\alpha$ is not known a priori, and it is reasonable to consider options alternative to static allocation of the resources.

\subsubsection{Four-step access}
\label{sec:four-step}
A four-step access procedure consists of the following steps:
\begin{enumerate}
\item The device sends a transmission request. The probability of error at the BS for this message is $\epsilon_R$.
\item The device waits to receive access grant denoting the reserved resources, error probability $\epsilon_G$. 
\item The device sends data in the uplink. Its error probability is $\epsilon_D$.
\item The devices waits to receive an ACK, whose error probability is $\epsilon_A$.
\end{enumerate}
The overall probability of error for stochastic arrivals in a four-step procedure is
\begin{equation}
\label{eq:ErrorURLLC4steps}
\epsilon_{\textrm{sto}4} = 1 - (1 - \epsilon_\text{sync})(1-\epsilon_R)(1-\epsilon_G)(1-\epsilon_D)(1-\epsilon_A).
\end{equation}
In contrast to~\eqref{eq:prealloc}, the term $\epsilon_\text{sync}$ here refers to the initial synchronization and not the absolute time synchronization, as the latter is typically not required for URLLC services with stochastic arrivals due to their less stringent requirements.
This initial synchronization is established through reception of the synchronization information via downlink broadcast channels. It contains carrier and frequency synchronization that will be exploited by the device for the subsequent uplink transmissions.\footnote{One could thus argue that the four-step procedure actually involves five steps, where the first step is the one related to successful reception of downlink broadcast information. We also note that similar considerations apply to the rest of the considered access procedures.}

For a given target overall error probability $\epsilon_\text{target}$, meeting the requirement $\epsilon_{\textrm{sto}4}\leq \epsilon_\text{target}$ is more difficult than meeting $\epsilon_{\textrm{det}}\leq \epsilon_\text{target}$, as~\eqref{eq:ErrorURLLC4steps} contains contribution from more steps than~\eqref{eq:prealloc}.
However, four-step access makes sense if $\Pr [ \alpha = 1 ] \ll 1$, as it aims to support only the active devices and thus offers an overall better use of resources then the one with the static allocation.

The error probabilities in four-step access $\epsilon_i$, $i \in \{R, G, D, A \}$ depend on the choice of the transmission parameters, but also on some other aspects of the access procedure operation.
Specifically, transmission requests are typically sent using a contention procedure among active devices; an important example of such access networking can be found in LTE.
In this case, sending of a request is subject to the interference $w$ that includes the request of its contenders.
Thus, special provisions should be made to keep the value of $\epsilon_R$ low.
A potential approach is to increase the number of resources for contention, and thus statistically decrease the interference generated by the other contenders.
However, the standard contention algorithms, like slotted ALOHA, are rather resource-inefficient, especially if target collision probabilities are low.
This calls for the application of contention procedures that are better suited to deal with interference, e.g. through use of successive interference cancellation~\cite{PSLP2014}, multi-packet reception~\cite{GSP2018,LLYPSC2018}, etc.

Another approach to keep $\epsilon_R$ low is to statically allocate the resources for sending the transmission request, no matter whether the user is active ($\alpha = 1$) or not ($\alpha = 0$).
This approach is similar to the one considered in Section~\ref{sec:deterministic}, also suffering from inflexibility.
However, if the amount of resources required to send a request is much smaller than the amount of resources required for the data transmission, this approach may attain the desirable efficiency.

Note that the successful reception of the request enables the BS to estimate the timing offset of the device and consequently instruct the device via grant message to compensate this offset in its subsequent data transmission. In other words, the estimation of the timing offset and its subsequent compensation effectively play the role of \emph{frame synchronization}. In this respect, access requests typically include metadata that fosters estimation of the timing offset.\footnote{In LTE access networking, this metadata is an Zadoff-Chu sequence; Zadoff-Chu sequences feature favorable auto- and cross-correlation properties.}

The probability of not receiving an access grant, $\epsilon_G$, depends on the fact whether BS has sufficient data resources to grant to all requests, as well as the correct reception of the access grant packet.
This kind of error can be influenced by the scheduling and resource allocation policy to other users and services.

For the sake of completeness, we note that the access procedures for the initial connection establishment in mobile cellular standards actually involve more than 4 steps before the data transmission can take place.
For instance, in LTE connection-establishment~\cite{TS36.321}, a device that wants to establish a connection and send data has first to successfully send a series of uplink messages with metadata\footnote{Specifically, ten messages with metadata~\cite{madueno2016assessment}.} used for timing-offset estimation, device's identification and notification of the reason for connection establishment, security context establishment, etc.
If a device is only sporadically active, the connections it establishes will become released.
This implies that sporadically active devices will have to undergo the connection-establishment procedure each time they have to send data. 
Obviously, this represents a huge challenge from both latency and reliability perspectives, a topic that has attracted a lot of attention in the recent literature on efficient support of machine-type communications in cellular access~\cite{madueno2016assessment}.

\subsubsection{Three-step access}
In the three-step access, sending of a request is skipped and the BS sends directly the access grant to poll the device.
In this case, the value of the activity indicator $\alpha$ becomes set to 1, no matter whether the device has experienced a new packet arrival or not.
This mode of operation makes sense for services in which the devices are polled when their data becomes needed, or in which the BS can accurately predict when the device wants to send data, i.e, predict when $\alpha$ has changed from $0$ to $1$.
Note that an inaccurate prediction results in either resource waste, as resources are allocated when no URLLC message is pending, or an outage, when the resources are not allocated and the message expires until the next transmission opportunity.

The overall probability of error the three-step procedure is:
\begin{equation}
\label{eq:ErrorURLLC3steps}
\epsilon_{\textrm{sto}3} = 1 - (1 - \epsilon_\text{sync})(1-\epsilon_G)(1-\epsilon_D)(1-\epsilon_A).
\end{equation}

In contrast to the four-step access and due to the lack of the timing offset estimation, a correct reception of the data transmission in the three-step access is more challenging.
Specifically, the BS has to detect where is the start of the data transmission in the dedicated resources, i.e., to acquire frame synchronization directly on the data transmission itself.
We turn to this problem in more details in Section~\ref{sec:frame-sync}.


\subsubsection{Grant-free access}
The 3-step procedure can be further decreased to a 2-step procedure, termed \emph{grant-free access}, where the transmission of the grant by the BS is skipped. The first transmission is carried out by the device and it contains the actual data that should be sent during the access procedure.
The probability of error in this case is:
\begin{equation}
\label{eq:ErrorURLLC2steps}
\epsilon_{\textrm{sto}2}= 1 - (1 - \epsilon_\text{sync})(1-\epsilon_D)(1-\epsilon_A).
\end{equation}
Although similar in form, the semantic difference between \eqref{eq:ErrorURLLC2steps} and \eqref{eq:prealloc} is substantial. In case of grant-free access, the data transmission 
is by default contention-based and subject to potential interference from data transmissions of other devices.
The modest performance of the standard contention algorithms necessitates consideration of more advanced solutions that are able to deal with interference, see~\cite{PSLP2014,GSP2018,LLYPSC2018}.
Moreover, similarly to the three-step access, the BS has to acquire frame synchronization.
Nevertheless, there are at least two reasons to use grant-free access: (1) it decreases  latency and (2) if the URLLC packets are very short, then the overhead brought by the request/grant is very significant and impacts the system efficiency.

We also remark that the assumptions for grant-free access can be further relaxed by not assuming prior synchronization between the base station and the devices in its domain. A prominent example of such approach is the pure ALOHA.
However, when the reliability requirements are very stringent, this type of solutions is infeasible, in particular when the load/interference of the access is high.

\begin{figure}[t]
    \centering
    \includegraphics[width=0.8\linewidth]{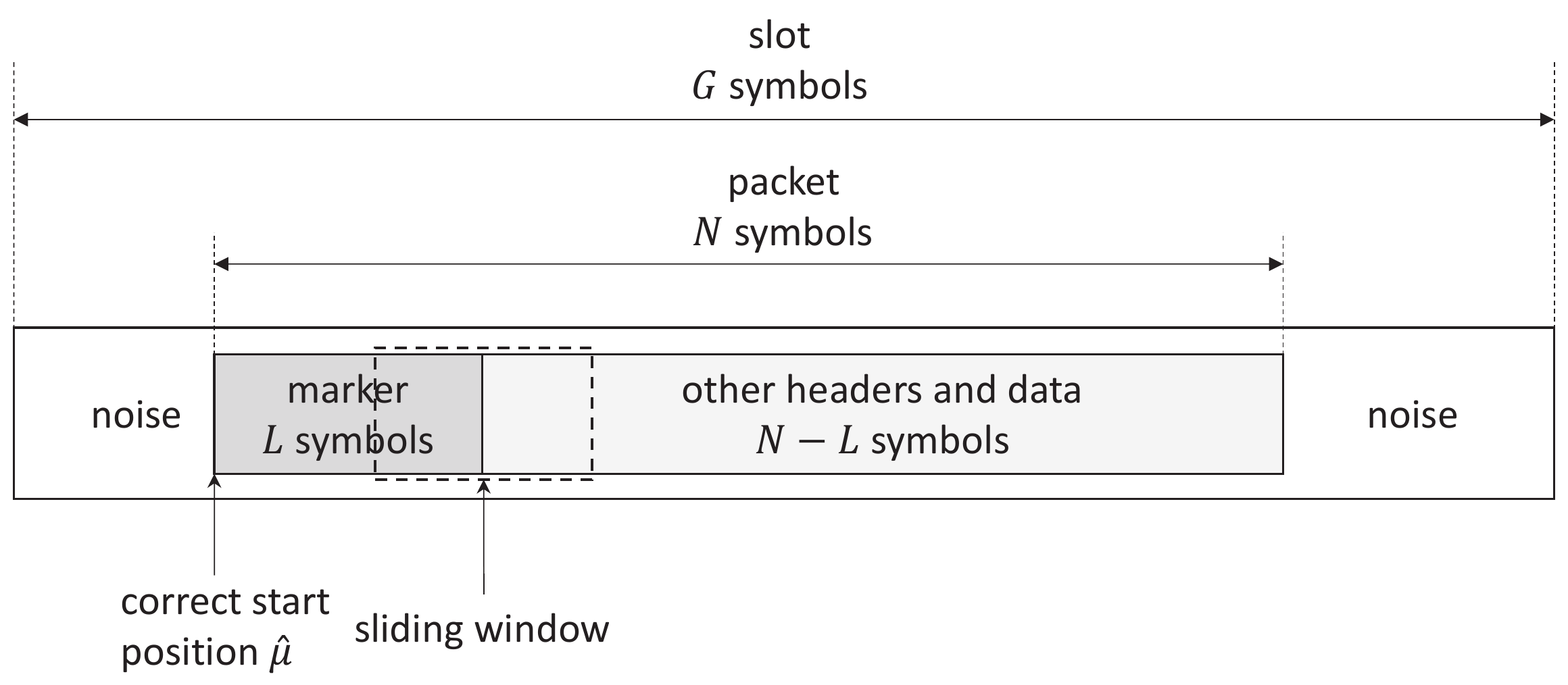}
    \caption{Frame synchronization: the receiver should detect the correct value of $\hat{\mu}$.}
    \label{fig:frame_sync}
\end{figure}

\subsection{Frame Synchronization}
\label{sec:frame-sync}

The task of the frame synchronization is to establish where is the start of the received packet. This is a critical ingredient of all access procedures discussed above and in this section we provide a brief discussion of the factors that affect its reliability, as well as an illustration of the required resources to support very high reliability.

A common approach in frame synchronization is to place a \emph{marker}, also known as \emph{synchronization sequence}, at the beginning of the packet, and the task of the receiver is to detect the marker in the observation made in the slot, see Fig.~\ref{fig:frame_sync}.
An alternative approach is to employ \emph{blind frame synchronization}, which exploits other forms of redundancy contained in the packet, notably the knowledge of the channel coding algorithm, in order to establish frame synchronization.
However, this approach suffers from higher complexity.

For the marker-based approach, the optimal detection algorithm depends on the channel model~\cite{M1972,R1994}.
In high signal-to-noise-ratio regimes, the optimal detection algorithm can be well approximated with the correlation between the locally generated marker and the subset of the received symbols that are placed in the sliding window. The length of this window is equal to length of the marker and it slides symbol-by-symbol through the slot, see Fig.~\ref{fig:frame_sync}.
Upon performing correlations for all window positions, the receiver selects the one with the highest correlation value.

If the impact of noise can be neglected, then the receiver may miss the correct start of the frame if the marker happens to be generated in the rest of the packet by chance.
To illustrate this, let us consider a high SNR scenario, in which BPSK modulation is used.
Further, assume that the marker length is $N_\text{m}$ bits, the total packet length is $N_\text{p}$ bits, and that the bit values of 0 and 1 in the rest of the packet are equiprobable, which is a reasonable assumption. Note that this is different from the AWGN channel discussed in Section~\ref{eq:URLLCpacket}, as here the input is set to be binary.
The upper bound on the probability of correct frame synchronization can be computed as
\begin{align}
\label{eq:UB}
P_\text{UB} = \sum_{i} \frac{1}{ i + 1 } \Pr \{ C = i \}
\end{align}
where $C$ is a random variable denoting the number of times that the marker is randomly reproduced by the packet symbols.
In the expression for $P_\text{UB}$ we have neglected the impact of the noise samples surrounding the packet, see Fig.~\ref{fig:frame_sync}.
$P_\text{UB}$ can be computed using the method presented in~\cite{SB2012}.

\begin{figure}[t]
    \centering
    \includegraphics[width=\linewidth]{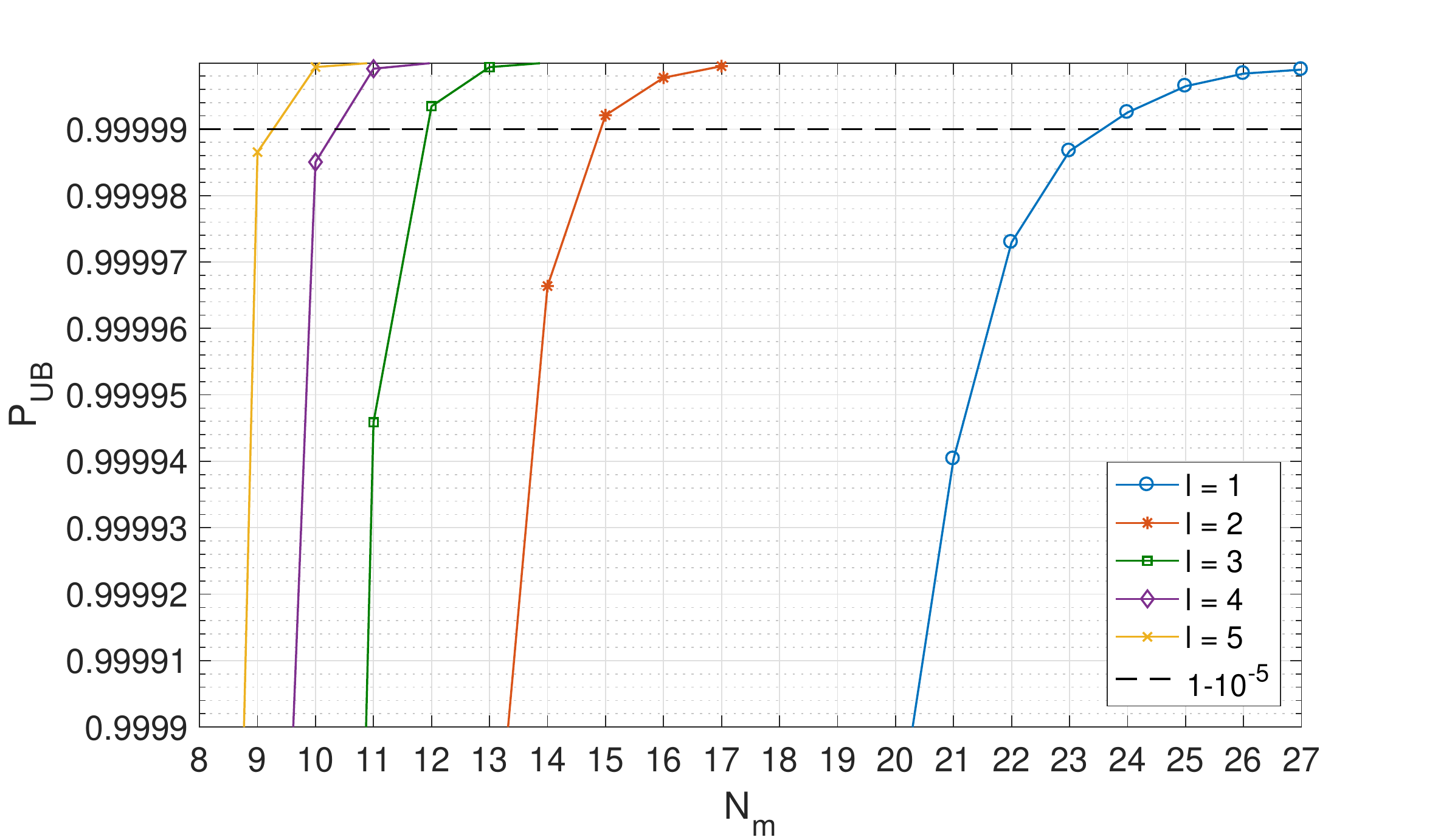}
    \caption{Upper bound on the probability of correct frame synchronization as function of marker length $N_\text{m}$ bits and of list length $l$, where packet length is $N_\text{p} = 256 + N_\text{m}$ bits.}
    \label{fig:upper-bound}
\end{figure}

Fig.~\ref{fig:upper-bound} shows $P_\text{UB}$ (the line marked with circles) as function of the marker length $N_\text{m}$, assuming that $N_\text{p} = N_\text{m} + 256$~bits, i.e. the packet length is $32$ bytes, not taking into account the marker.
The presented results are obtained using the marker patterns that follow standard design guidelines~\cite{SJ1988,GG2009}.
If one wants to achieve the correct frame-synchronization performance of $1 - 10^{-5}$, thus matching the standard URLLC reliability requirement~\cite{TR38.913}, the marker length should be larger than $24$ bits, even in the high SNR regime. 

A compromise between complexity and performance can be achieved using a two-stage list-based synchronizer.
The output of the first stage are $l$ positions that are most probable to contain the marker, obtained using the marker detection algorithm.
In the second stage, one of these $l$ positions is selected using a metric that exploits the knowledge of the channel coding algorithm.
Analogously to \eqref{eq:UB}, we can calculate an upper bound of the correct frame synchronization related to the probability that the output of the first stage contains the correct frame start position:
\begin{align}
\label{eq:UB-list}
P_\text{UB}(l) = \Pr \{ C < l \} + \sum_{i\geq l} \frac{l}{ i + 1 } \Pr \{ C = i \}.
\end{align}
Fig.~\ref{fig:upper-bound} presents the results for increasing values of $l$, also obtained via the method from~\cite{SB2012}.
It can be seen that, for a fixed value in $P_\text{UB}$, the increase in $l$ allows to decrease $N_\text{m}$.

Finally, we also mention that frame synchronization can be achieved by using markers whose symbols belong to different alphabets in comparison to the symbols carrying data in the rest of the packet.
For example, an option could be to use Zadoff-Chu sequences as markers, cf.~\cite{APMS2017}.
Nevertheless, the results concerning the marker lengths and/or receiver complexity, presented above, also hold qualitatively in those cases.

\section{URLLC in Massive Multi-Antenna Systems}
\label{sec:MassiveMIMO}

\subsection{The Benefits of Massive Multi-Antenna Systems for URLLC}

Multiple antennas at the base station (BS) or terminals of a wireless network provide efficient mechanisms at the physical layer to ensure reliable and low latency communications.  They offer a powerful complement to the higher layer methods described in this paper. 
This section  focuses on massive antenna systems, characterized by a very large number of antennas at the BS and, possibly at the terminals, at high frequency bands, that have clearly emerged as a major enabler towards the creation of 5G wireless networks~{\cite{five_disruptive_tech_5g}}. They are largely viewed as essential in magnifying the data rates and/or increasing the number of broadband users that can be simultaneously multiplexed within the same bandwidth. However, they are also fundamental tools in building the two other 5G services, i.e. massive machine type communications {\cite{massivemimo_massive_connect}} and URLLC \cite{feasibility_large_arrays_urllc}, \cite{urllc_mmwave_massivemimo}.

The benefits of massive antenna systems lie in their ability to create a very large number of spatial Degrees-of-Freedom (DoF), which determine 
the following remarkable properties that are beneficial for URLLC:
\begin{enumerate}
    \item \emph{High SNR links.} This property is due to the array gain. 
    \item \emph{Quasi-deterministic links, practically immune to fast fading.}  This property is rather specific to systems operating below 6~GHz in a rich scattering environment~{\cite{favorable_propag}}. It is a result of the \emph{channel hardening} phenomenon. Along with the first property, it relaxes the need for strong coding schemes, hence maintaining high reliability for short packets. This can dramatically reduce the need for retransmissions. 
    \item \emph{High capability for spatial division multiplexing.} In a multi-user system, this property can be exploited to improve the latency incurred due to multiple access, as multiple users can exchange data simultaneously. However, it should be noted that the multi-antenna processing employed to separate the users might induce additional computational delay~{\cite{massive_MIMO_tactile_internet}}. 
\end{enumerate}

This section is dedicated to the exploitation of multiple antennas at the transmitter and receiver to support URLLC. 
At first, we need to establish the fact that the acquisition of the instantaneous CSI is one of the most severe limitations with respect to URLLC when exploiting multiple antennas; see Section~\ref{sec:RelatedWork}. This is because the CSI acquisition is a major protocol step in massive MIMO, impacting both the reliability and the latency. Taking this into account, we devise beamforming methods that rely mostly on the structure of the channel, that is, the direction of the propagation path. The information about small scale fading is exploited as little as possible. As the structure of the channel  varies on a large scale basis, its acquisition is  more robust to device mobility. 

It should be noted that the basic idea of using the singular vectors of the channel~\cite{Boccardi09} or the structure of the channel~\cite{JSDM13, Chowdhury17} (singular vector of the covariance matrix or steering vectors)  to build multi-user transceivers is not new. Here we show that this basic idea creates a good basis to build URLLC transmission 
schemes.


\subsection{Channel Structure}

\begin{figure}[t]
    \centering
    \includegraphics[width=\linewidth]{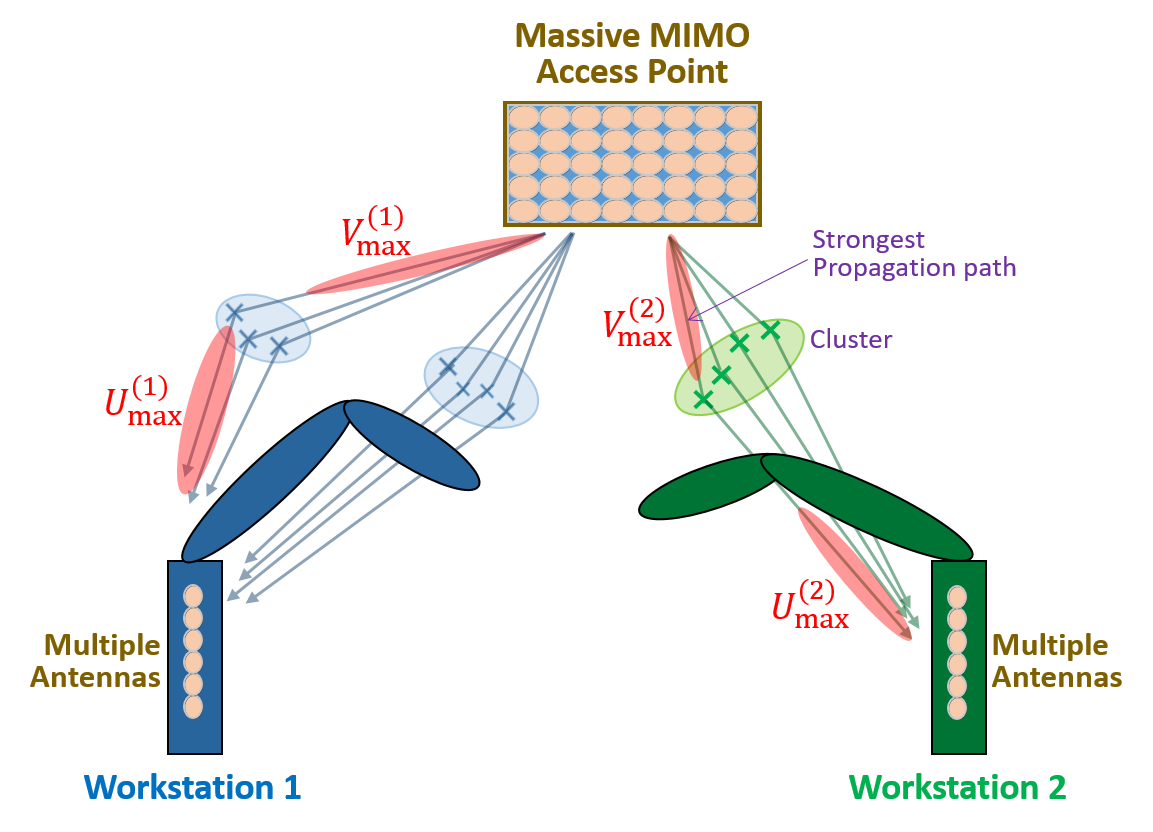}
    \caption{Factory scenario where a massive MIMO access points serves multiple terminals (workstations).}
    \label{fig:MIMO_Factory}
\end{figure}

To illustrate the main concepts of this section, we assume a factory-type environment as pictured in 
Fig.~\ref{fig:MIMO_Factory}. 
An access point is equipped with an array that consists of a very large number of antennas while the terminals (workstations) are equipped with one or possible small number of antennas.

Furthermore, we consider the simplified case of two terminals, each receiving a single stream of data from the access point. The results can be easily extended to the general case. 
We adopt a cluster-based channel model where each cluster is characterized by a group of localized propagation paths defined by their direction of departure and their direction of arrival. Each propagation path is affected independently by an attenuation factor that follows a certain distribution. 
The channel from the access point to  terminal $k$ is described as the sum of the propagation paths over all the clusters (in the sum, we make no distinction between clusters):
\begin{eqnarray}
\bfH^{(k)} =  \sum_{i=1}^{N_P^{(k)}} \alpha_i^{(k)} \bfs_{i,\text{rx}}^{(k)} \bfs_{i,\text{tx}}^{(k)H}.
\end{eqnarray}
$N_P^{(k)}$ is the total number of paths. 
$\bfs_{i,\text{tx}}^{(k)}$ and $\bfs_{i,\text{rx}}^{(k)}$ are the normalized steering vectors which characterize respectively the direction of departure from the BS and direction of arrival to the terminal.
When a terminal is equipped with a single antenna, we have $\bfs_{i,\text{tx}}^{(k)} =1$.
The direction of the propagation paths correspond to long-term statistics, meaning that for a localized movement of the terminal the directions remain unchanged, whereas the coefficients 
$\{\alpha_i^{(k)}\}$ corresponds to small scale fading and vary for small  movements. 

\subsection{Covariance-based Design}

In order to promote reliability and low latency, 
the general purpose of the beamforming design is to rely as much as possible on the structure of the channel (i.e. the propagation path) and as less as possible on the small scale fading properties, while still benefiting from the properties brought by the massive number of antennas.  
This, in general, is a non-trivial task as those properties are brought by coherent combining of the signals from each antenna, while this combining depend on the small-scale fading. 

We adopt a design based on the covariance matrix of the signal of each terminal at the transmitter and receiver. Those covariance matrices reflect the structural properties of the channel. 

The  singular value decomposition of the covariance matrix at the transmitter for terminal k is: 
\begin{eqnarray}
\bfR^{(k)}_{\text{tx}} 
= {\bfV}^{(k)} \Lambda^{(k)} {\bfV}^{(k)H}.   \label{eq:cov2}
\end{eqnarray}
The columns of ${\bfV}^{(k)}$ comprise the singular vectors denoted as $\bfv_{i}^{(k)}$ and  $\Lambda^{(k)}$ is a diagonal matrix grouping the non zero singular values. 
Likewise, we write the covariance matrix at  terminal $k$ as:
\begin{eqnarray}
\bfR^{(k)}_{\text{rx}} = 
{\bfU}^{(k)} \Lambda^{(k)} {\bfU}^{(k)H}.   \label{eq:Rxcov2}
\end{eqnarray}
The singular vectors of $\bfR^{(k)}_{\text{rx}}$ are denoted as $\bfu_{i}^{(k)}$. 
The singular vectors associated to the maximal singular value  $\bfR^{(k)}_{\text{tx}}$ and 
 $\bfR^{(k)}_{\text{rx}}$ are
$\bfv_{\max}^{(k)}$ and $\bfu_{\max}^{(k)}$.

\subsection{Transceiver Structures}
\label{sec:ZF}

We now examine zero-forcing beamforming designs. They are based on the following principles:

1) {\bf Inter-terminal properties}: The inter-terminal interference can be removed based solely on the singular vectors of the covariance matrix of the interfering terminals defining their signal subspace. 
Interference is eliminated by projecting the transmitted signal into the space orthogonal to signal subspace of the interferers.    
This is advantageous in URLLC as this operation does not depend on instantaneous CSI. 

2) {\bf Intra-terminal properties}: Once the inter-terminal interference is removed, transmission to a single terminal might exploit several levels of CSI knowledge at the transmitter.

A general form of the zero-forcing precoder for terminal 1 is as: 
\begin{equation}
\bfF_{\text{ZF}}^{(1)}  = 
\underbrace{P^{\perp}_{{\cal V}^{(2)}}}_{\text{Term 1}}  
\;\;
\underbrace{{\bar{\bfV}}^{(1)}_{~}}_{\text{Term 2}}
\;\;
\underbrace{\bfw_{~}}_{\text{Term 3} }
\label{eq:ZF}
\end{equation}

\noindent
{\bf  Term 1.}
The first term forces the precoded signal to lie in the signal subspace orthogonal to terminal 2. 
${\cal V}^{(2)}$ contains the  singular vectors of the covariance matrix at the transmitter for terminal 2 associated to non-zero singular values. 

\noindent
{\bf  Term 2.}
The columns of the matrix in the second term defines the subspace of the transmit covariance matrix of terminal 1 where the signal of interest lies. 
${\bar{\bfV}}^{(1)}$ contains  the singular vectors   of the covariance matrix  associated to non-zero singular values. 

\noindent
{\bf  Term 3.}
Vector $\bfw$ defines a linear combination of the columns of  ${\bar{\bfV}}^{(1)}$. In general, this is a coherent operation requiring the knowledge of the channel projection 
onto the columns of  ${\bar{\bfV}}^{(1)}$. 

Note that term 1 and term 2 depend only on the long term statistics of the channel. 

%
 

\subsection{Beamforming Methods}

We test the following transceiver structures that are classified by decreasing level of instantaneous CSI they exploit:
\begin{itemize}
\item {\bf Interference free}: as a performance upper bound, we plot the case where the inter-terminal interference is ignored. 

\item  {\bf  All SV - Coh}: transceiver according to equation (\ref{eq:ZF}) where 
all effective SVs are considered and coherent combining is performed. Information about the instantaneous CSI is needed.  
\item {\bf Strongest SV - Inst}: transceiver according to equation (\ref{eq:ZF}) with ${\cal V}^{(2)} = \bfU^{(2)}$ and
${\bar V}^{(1)} = \bfv^{(1)}_{I,\max}$. 
This strategy necessitates partial instantaneous CSI at the transmitter. 
Assuming that the receiver applies $\bfu^{(1)}_{\max}$, the transmitter estimates the projection of $\bfH^{(1)}$ 
into the singular vectors  ${\bar V}^{(1)}$ and selects  the strongest one, denoted as $\bfv^{(1)}_{I,\max}$.

\item  {\bf All SV - NCoh}: transceiver according to equation (\ref{eq:ZF}) where  all SV are considered. 
Transmission across the singular vectors is performed non-coherently. 
The transmit power along singular vector $\bfv^{(1)}_{i}$ is $\lambda^{(1)}_{i}$.

\item {\bf  Strongest SV - Av}:  transceiver according to equation (\ref{eq:ZF}) 
with ${\cal V}^{(2)} = \bfU^{(2)}$
and
${\bar{\bfV}}^{(1)} = \bfv_{\max}^{(1)}$.

\end{itemize}

For the methods relying on the whole set of  singular vectors ("All SV"), the receiver estimates the aggregate channel matrix 
$\bfH^{(1)} \bfF_{\text{ZF}}^{(1)} $ and matched filtering is applied. 
For the other methods, the receiver applies the filter matched to $\bfv_{\max}^{(1)}$ and only requires the estimation of the projection of the aggregate channel on  
$\bfv_{\max}^{(1)}$. 

\subsection{Numerical Evaluations}

\begin{figure}[t]
    \centering
    \includegraphics[width=\linewidth]{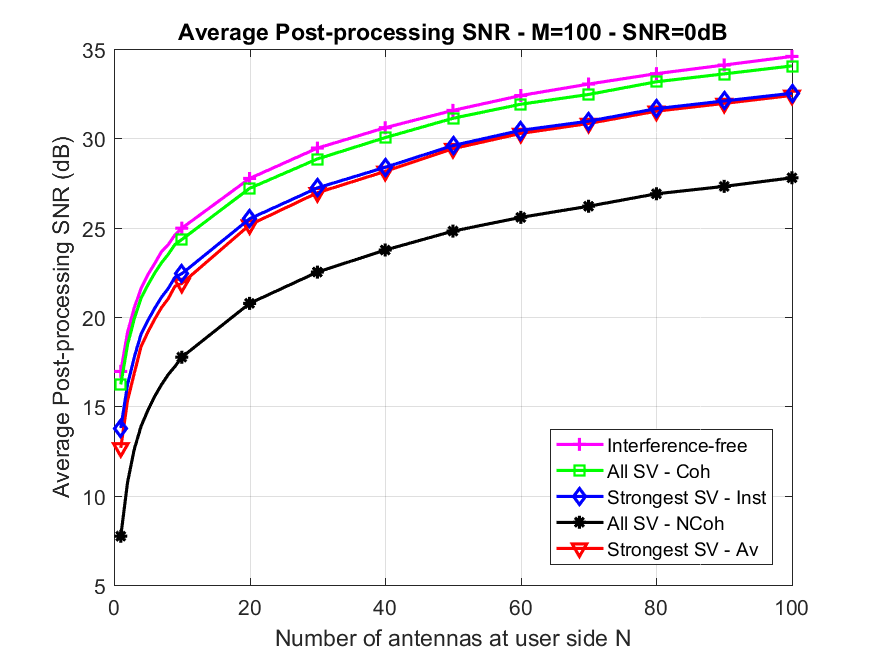}
    \caption{SINR in a 2-user scenario vs number of antennas at the terminals, $\rho=0$dB. }
    \label{fig:SINR}
\end{figure}

\begin{figure}[t]
    \centering
    \includegraphics[width=\linewidth]{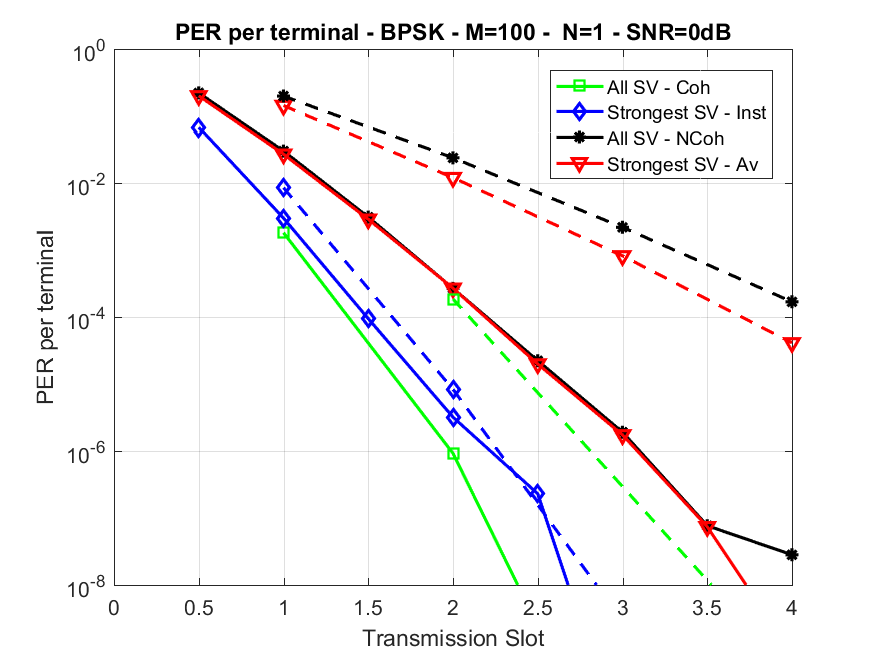}
    \includegraphics[width=\linewidth]{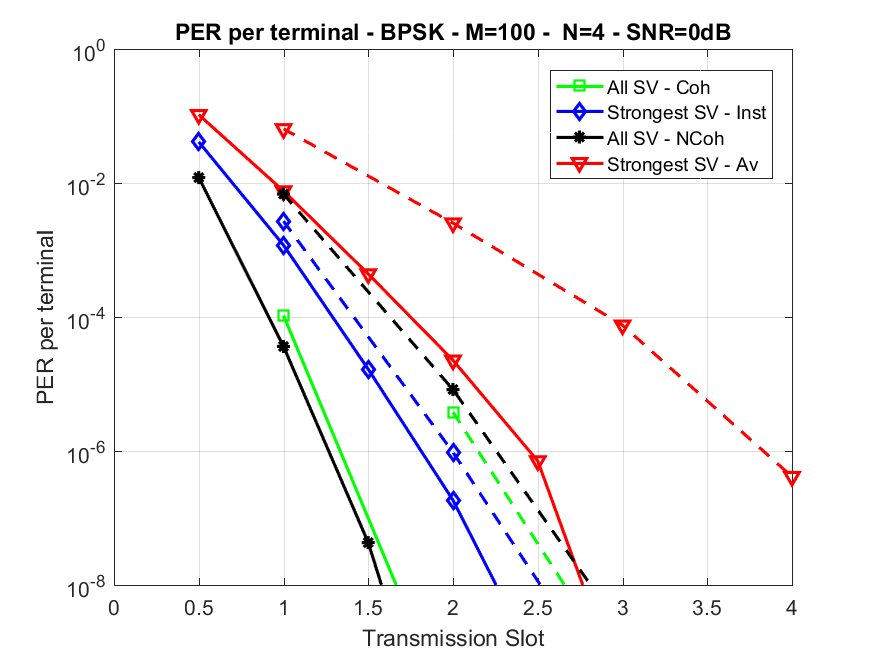}
    \caption{Packet error rate in a 2-user scenario vs transmission slot, $\rho=0$dB.  Users are spatially multiplexed (solid lines) and time multiplexed (dashed lines). }
    \label{fig:PER}
\end{figure}

Fig.~\ref{fig:SINR} and  Fig.~\ref{fig:PER}  
display the post-processing SINR and the Packet Error Rate (PER) associated to the different transceiver structures. 
The total number of antennas at the access point is $M=100$ and the SNR  is defined as $\rho=P/\sigma_n^2$ where $P$ is the total transmit power and $\sigma_n^2$ is the variance of the noise at each receiving antenna. We normalize 
$\bfF_{\text{ZF}}$ in 
 (\ref{eq:ZF}) so that the transmit power is divided equally among the users. 
The multipaths in a single cluster are assigned different delays with an exponential decay that is up to 20dB. 

Fig.~\ref{fig:SINR} shows the post-processing SINR as a function of the number of antennas at the terminal side. 
As expected, we observe a gap between the methods exploiting full CSI and the methods based on second-order statistics or partial CSI. There is little differentiation for the latter methods. 

Fig.~\ref{fig:PER}  displays the Packet Error Rate (PER) as a function of the transmission slot  for the case of a single antenna per user. 
The bandwidth is normalized, such that the number of channel uses directly reflects the delay.
The payload is composed of 100~bits drawn from a BPSK modulation, hence we transmit 1~bit per channel use. 
We assume that the duration of the training for the coherent transmission techniques is twice as much as for the methods based on partial and no instantaneous CSI. A transmission slot is defined as the duration to send a packet (payload and overhead) using coherent transmission. Within a transmission slot, two packets can be sent using non-coherent transmission. 
The case where the users are multiplexed in space (solid lines) or in time (dotted lines) is shown. 

For the selected simulation parameters, the following observations can be highlighted: 

\begin{itemize}

\item  The general tendency is that performance gets better with an increased exploitation level about the channel at the transmitter. 

\item There is a notable exception when the terminals are equipped with multiple antennas and receive diversity is exploited. 
In Fig.~\ref{fig:PER}, for $N=4$,  the non-coherent strategy ("All SV - NCoh") performs the best. 
Hence, from a BER perspective, it is preferable to transmit the signal in a non-coherent fashion  along each singular vector.  
The non-coherent transmission is compensate for by a receive coherent processing by multiple antennas that  allows the extraction of diversity. 

\item Depending on the level of CSI exploited at the transmitter, space multiplexing is not always favourable. 
\end{itemize}

%
%
%
%
%
%
%

%

\section{Multi-Connectivity and Interface Diversity}
\label{sec:IFD}

The mobile devices today have multiple radio interfaces and it is likely that many of the future devices will have that as well. This is also an indicator that the 5G radio interfaces will be deployed along with other radio interfaces. From the perspective of URLLC, the existence of multiple interfaces offers an additional degree of diversity that can be used to achieve the stringent latency-reliability requirements. This is commonly known as \emph{multi-connectivity}~\cite{SWDBK2018}, while here use the terms \emph{link diversity} or \emph{interface diversity} in order to emphasize the diversity role played by the availability of multiple different communication interfaces. 

The idea of using multiple links or interfaces simultaneously is fairly natural and it has already emergent in some settings. In the context of 3GPP systems, LTE has supported Multi-Connectivity through Carrier Aggregation (CA) and Dual Connectivity (DC) since rel. 10 and 12, respectively. However, in this case the objective is throughput enhancement. Recently, in Rel. 15, Packet Duplication was introduced by 3GPP to boost reliability~\cite{SWDBK2018}. The data packet is duplicated on PDCP and transmitted on independent channels, either from the same eNB on different carriers via CA or from different eNBs using DC.

\begin{figure}[t!]
    \centering
    \includegraphics[width=0.8\linewidth]{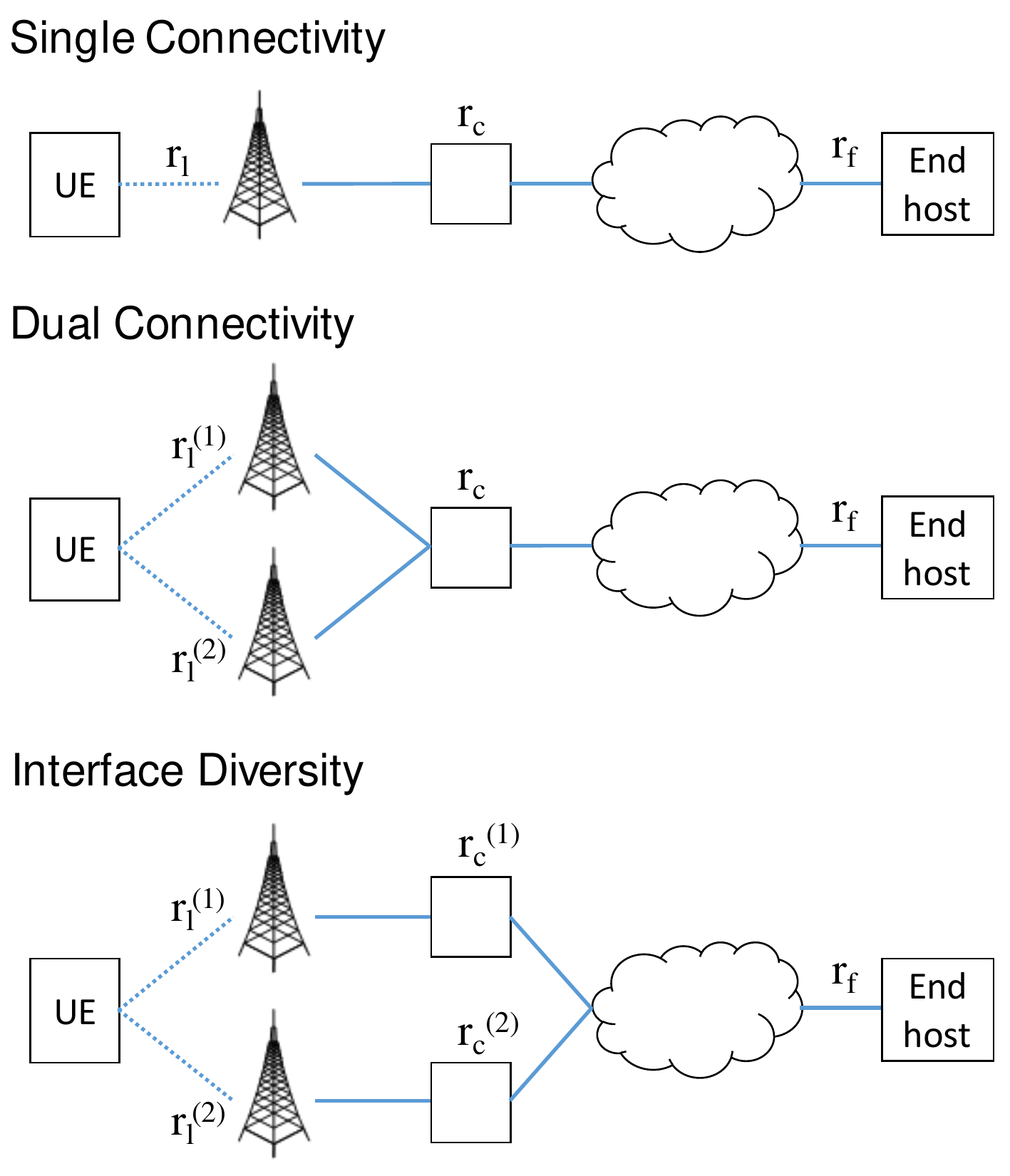}
    \caption{Dual Connectivity and Interface Diversity architectures.}
    \label{fig:reliability_architectures}
\end{figure}

Packet Duplication in Multi-Connectivity architectures are excellent for mitigating losses due to fading and interference on individual links or temporary scarcity of air interface resources. Nevertheless, the reliability of the end-to-end connectivity relies on the correct functioning of an infrastructure and core network, often belonging to a single operator. While infrastructure and core networks are based on redundant solutions, they are still subject to single Point-of-Failure (PoF), e.g. through equipment misconfiguration.
This reliance on a single network infrastructure can be mitigated by providing diversity not only at a link level, but also a a level of a comunication interface or a path, as illustrated in Fig.~\ref{fig:reliability_architectures}. 

The concept of Interface Diversity (IFD) was studied in~\cite{nielsen2017ultra}. Interface Diversity provides an independent path from the UE to the internet (cloud), by the use of a different wireless technology and/or a different mobile network operator. That is, IFD can be obtained by equipping a device with, for example, LTE/5G and Wi-Fi interfaces, or LTE/5G interfaces with SIM cards from two physically independent mobile network operators. The key benefit of IFD is that there is no dependency on a single point of failure in the access network part. However, IFD requires that source and destination devices are configured to duplicate packets and handle multiple received copies, respectively. In comparison, dual connectivity is transparent to the source and destination devices, above the MAC layer. It should also be noted that Packet Duplication is the simplest instance of IFD, in which packet replicas are sent over different interfaces; more advanced IFD solutions involve various types of data segmentation and packet-level coding.

\subsection{Reliability Model and Numerical Illustration}
Assuming link/component reliabilities as indicated in the figure, we can use a series/parallel systems analogy from reliability engineering to express the end-to-end reliability of the architectures as:
\begin{align}
    R_\text{single} &= r_\text{l} r_\text{c} r_\text{f} \\
    R_\text{DC} &= \left(1-\prod_{i=1}^N(1-r_\text{l}^{(i)})\right) r_\text{c} r_\text{f} \\
    R_\text{IFD} &= \left(1-\prod_{i=1}^N(1-r_\text{l}^{(i)} r_\text{c}^{(i)})\right) r_\text{f},\label{eq:ifd}
\end{align}
where $R_\text{single}$, $R_\text{DC}$ and $R_\text{IFD}$ are for for a single link, $N$-link Dual Connectivity (DC) and $N$-interfaces Interface Diversity (IFD). $r_\text{l}^{(i)}$ and $r_\text{c}^{(i)}$ refer to the reliability of the $i$th link/interface or core network, respectively. A key assumption for eq. \eqref{eq:ifd} is that the considered interfaces are uncorrelated in the sense that failures are occurring independently. This can be ensured in practice by using different mobile networks that do not share physical infrastructure.

Let us initially consider the two-link/interface instances sketched in Fig. \ref{fig:reliability_architectures}. The assumed default parameters are given in Table \ref{tab:reliability_values}.

\begin{table}[h]
    \centering
        \begin{tabular}{ccc} \toprule
         & LTE/5G & Wi-Fi \\ \cmidrule{2-3}
        $r_\text{l}$ & 0.99 & 0.9 \\
        $r_\text{c}$ & 0.999 & 0.99 \\
        $r_\text{f}$ & 0.9999 & 0.9999 \\ \bottomrule
        \end{tabular}
    \caption{Assumed default reliability parameters.}
    \label{tab:reliability_values}
\end{table}

\begin{figure}[htb]
    \centering
    \includegraphics[width=0.9\linewidth]{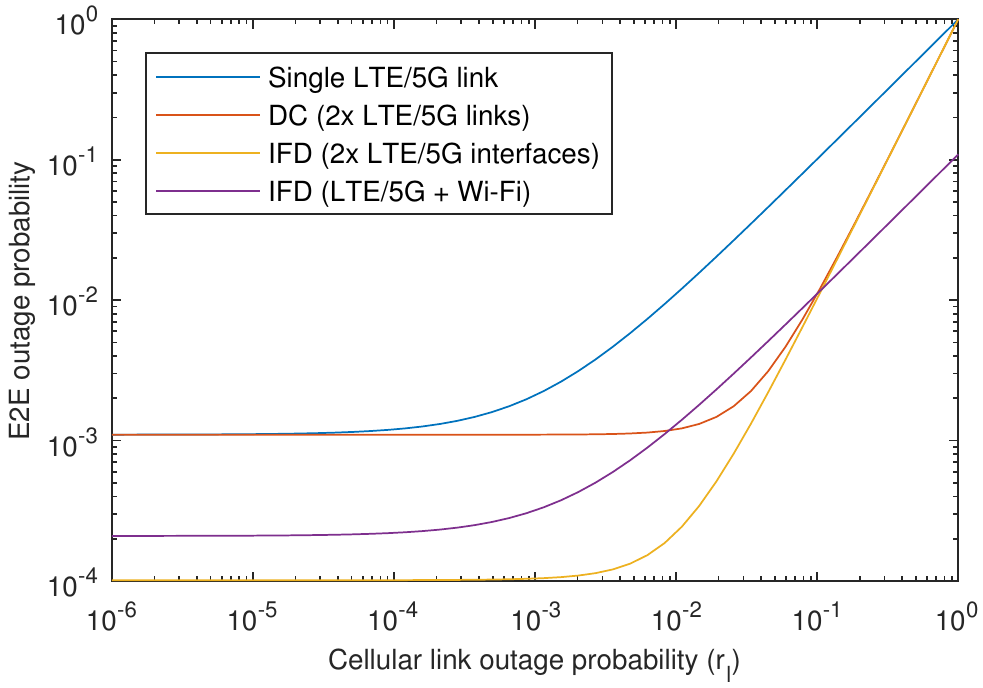}
    \caption{End-to-end outage probability for varying cellular link outage. Note that the assumed Wi-Fi link reliability is $r_{\text{l}_2}=0.9$.}
    \label{fig:outage_links_2-links}
\end{figure}

\begin{figure}[htb]
    \centering
    \includegraphics[width=0.9\linewidth]{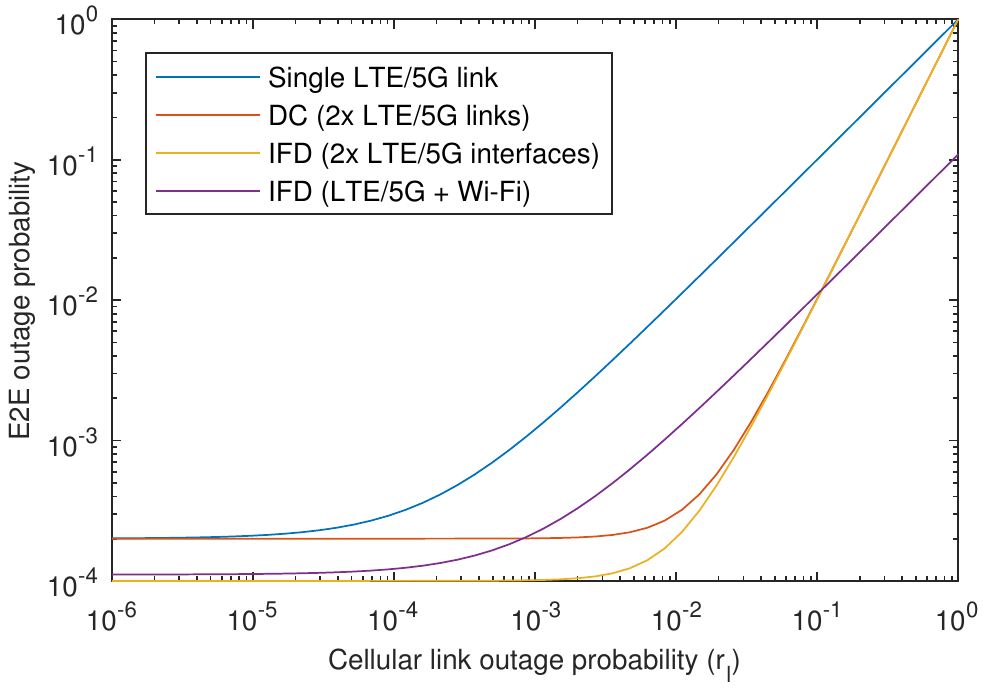}
    \caption{End-to-end outage probability for varying cellular link outage with $r_\text{c}=0.9999$. Note that the assumed Wi-Fi link reliability is $r_{\text{l}_2}=0.9$.}
    \label{fig:outage_links_2-links_UR-core}
\end{figure}

Fig. \ref{fig:outage_links_2-links} shows the resulting end-to-end outage probability when subject to different cellular link outages. The results show that IFD using two independent networks is always superior or equal in outage compared to DC. Specifically, when the cellular links are good, i.e. outage below $10^{-2}$, the outage of IFD is for an order of magnitude better compared to DC. Even the alternative configuration where a LTE/5G is complemented by an inferior, but independent Wi-Fi connection, is outperforming DC for link outages below $10^{-2}$. Further, the plot reveals that DC is better than using a single link, unless when the link outage is very low, and the end-to-end outage is instead dominated by the core outage probability. In comparison, consider the plot in Fig. \ref{fig:outage_links_2-links_UR-core}, where the mobile network core is assumed to be more reliable. In that case, the difference between DC and IFD is almost negligible. The advantage compared to using just a single link is significant, especially for link outages between $10^{-3}$ and $10^{-2}$.

\section{Statistical Aspects of Ultra-Reliable Guarantees}
\label{sec:LTK}


In this section we investigate fundamental statistical questions related to ultra-reliability; namely, how can high reliability be assessed and measured. The ambitious reliability figures in URLLC only make sense when they are related to a statistical model of the context/environment in which the URLLC system is deployed. However, the statistical model of the wireless context for URLLC is not known \emph{a priori} and sets the stage of the methods of \emph{statistical machine learning}, through which one can estimate the statistical properties of the environment and offer reliability guarantees.

Here we consider these questions in a simple setting: we investigate the impact of limited transmitter-side channel knowledge on the reliability performance of the communication links.
Specifically, we consider a one-way communication link where the transmitter sends a packet to a receiver at rate $R$ over a narrowband wireless channel; the model of the received baseband signal at the transmitter is given by \eqref{eq:BasicCommModel}.
To isolate and study only the impact of channel uncertainty, in this section we ignore the effect of noise and interference and consider packet errors due to \emph{outage}, defined by the following event:
\begin{equation}\label{eq:error_event}
R>\text{log}_2(1+P),
\end{equation}
where $P$ denotes the received power; from model \eqref{eq:BasicCommModel} we have that $P = |h|^2$ where we have normalized the transmit power $|x|^2$ to unity. 

Throughout the section, we will assume that the true distribution of the channel is circularly-symmetric $h\sim\mathcal{CN}(0,\sigma^2)$, implying that the received signal is dominated by scattered diffuse components; hence, the received envelope $\sqrt{P} = |h|$ follows Rayleigh distribution and the received power $P$ follows exponential distribution with scale parameter $\theta = 2\sigma^2$ denoting the average channel power. 
Under the above assumption, the \emph{outage probability} at transmission rate $R$ is given by the cdf of the received power:
\begin{equation}\label{eq:outage_prob}
F(R) = 1 - e^{-\frac{2^{R}-1}{\theta}}.
\end{equation}
Thus, transmitting at a specific rate 
over a Rayleigh channel with average power $\theta$, yields a specific outage probability.
The goal of ultra-reliable communication is to choose the \emph{maximum} rate that meets a predetermined reliability criteria.
However, as discussed in the following paragraphs, designing the reliability criteria and finding the most favorable rate is strongly linked to how much the transmitter knows about the channel.

\subsection{Na\"{i}ve rate selection under channel uncertainty}

First, consider the benchmark case where the transmitter knows the channel perfectly; this implies that the transmitter knows the average channel power $\theta$ also perfectly.
Under such circumstances, the transmitter can easily determine the maximum rate as a function of $\theta$ at which an outage probability level of $\epsilon$ can be guaranteed with \emph{certainty}:
\begin{align}\label{eq:mk_ecapacity1}
R_{\epsilon}(\theta) & = \sup\left\{R>0:\text{Pr}(R>\text{log}_2(1+P))\leq\epsilon\right\}\\\label{eq:mk_ecapacity2}
& = \text{log}_2(1 - \theta\ln(1-\epsilon)).
\end{align}
$R_{\epsilon}(\theta)$ is also known as $\epsilon$-outage capacity.

Now, consider the situation in which the transmitter has no knowledge of the average power $\theta$.
Instead, it collects $n$ independent and noiseless power measurements $x_1,\hdots,x_n$ via \emph{training}. 
Having acquired $x_1,\hdots,x_n$, a simple but na\"{i}ve solution would be for the transmitter to compute the Maximum Likelihood (ML) estimate of $\theta$ by averaging $x_1,\hdots,x_n$, i.e.
\begin{equation}
    \hat{\boldsymbol{\theta}}_{\text{ml}}(x^n) = \frac{1}{n}\sum_{i=1}^n x^i,
\end{equation}
and plug the obtained estimate in \eqref{eq:mk_ecapacity2} to determine the transmission rate $R(x_1,\hdots,x_n)$, which now becomes a random variable.
According to \eqref{eq:outage_prob}, every $R$ yields specific outage probability; hence, the sequence of random variables $X^n$ induces a distribution over $F$ and the transmitter can no longer guarantee that the outage probability under transmission rate $R(X^n)$ will be equal of less than $\epsilon$.

\subsection{Probabilistic rate-selection framework: parametric channel models}

The above discussion shows that when the transmitter has limited knowledge of the channel, it can only guarantee the reliability \emph{probabilistically}.
Formally, this can be done by choosing the most favorable, i.e. the largest \emph{rate-selection function} $R(X^n)$ such that predetermined \emph{statistical reliability constraint} is satisfied.
Depending on the specific formulation of the constraint, the channel knowledge status at the transmitter and the actual statistics of the channel, finding the most favorable rate-selection function might be involved problem.
In the rest of the section, we will limit our discussion to the following somewhat heuristic but intuitive choice.
Specifically, given specific realization $x^n$ of $X^n$, the transmitter uses the following transmission rate:
\begin{equation}\label{eq:rate_selection}
R(x^n) = R_{\epsilon_n}\left(\hat{\boldsymbol{\theta}}_{\text{ml}}(x^n)\right),
\end{equation}
for some $\epsilon_n>0$; our aim is to find $\epsilon_n$ for each $n$ such that $R(x^n)$ is maximized under predefined reliability constraint.
The outage probability under transmission rate selected according to \eqref{eq:rate_selection} is still a random variable; however, selecting $\varepsilon_n$ according to specific reliability constraint effectively controls the amount of uncertainty in the outage probability.
Note that for $\epsilon_n=\epsilon$ we have the na\"{i}ve solution.
Finally, we expect that the transmission rate is consistent, i.e. as $n\rightarrow\infty$, $R(x^n)$ converges to the $\epsilon$-outage capacity.

Choosing the specific formulation of the statistical reliability constraint can be done in many ways.
Here, we will consider two approaches, described next.

\subsubsection{Average Reliability (AR)}

Consider the following constraint:
\begin{equation}\label{eq:const1}
\sup_{\theta}\text{Pr}[R(X^n)>\text{log}_2(1+Y)]\leq\epsilon,
\end{equation}
where the averaging is performed w.r.t. the joint distribution of $Y,X^n$.
By rewriting the above according to the rule for total probability, by first averaging over $Y$ and then averaging over $X^n$, we observe that \eqref{eq:const1} guarantees that the worst-case \emph{mean} of the outage probability $F(X^n)$ will remain below $\epsilon$.
Note that this does not guarantee that for some specific realization of $X^n$ the outage probability will not be larger than $\epsilon$.
Regarding potential use-cases, constraint \eqref{eq:const1} is suitable for dynamic environments and can be used when one wishes to optimize the rate of the system and provide reliability guarantees \emph{jointly} over the training and the transmission, prior to the actual training.

Using the Rayleigh-channel assumption and the rate-selection function \eqref{eq:rate_selection} and after few elementary computations, we obtain the following:
\begin{align}
& \sup_{\theta}\text{Pr}[R(X^n)>\text{log}_2(1+P)] = \\
& = 1 - \left(1 - \frac{\ln(1-\epsilon_n)}{n}\right)^{-n}.
\end{align}
The maximum value of $\epsilon_n$, satisfying \eqref{eq:const1} is given by:
\begin{align}\label{eq:varen_constI_rayleigh}
\varepsilon_{n} = 1 - e^{-n\left((1-\epsilon)^{-\frac{1}{n}} - 1\right)}.
\end{align}
Clearly, as $n\rightarrow\infty$, $\varepsilon_n\rightarrow\epsilon$.

\begin{figure*}[t]
\centering
\subfloat[Average reliability constraint]{\includegraphics[scale=0.61]{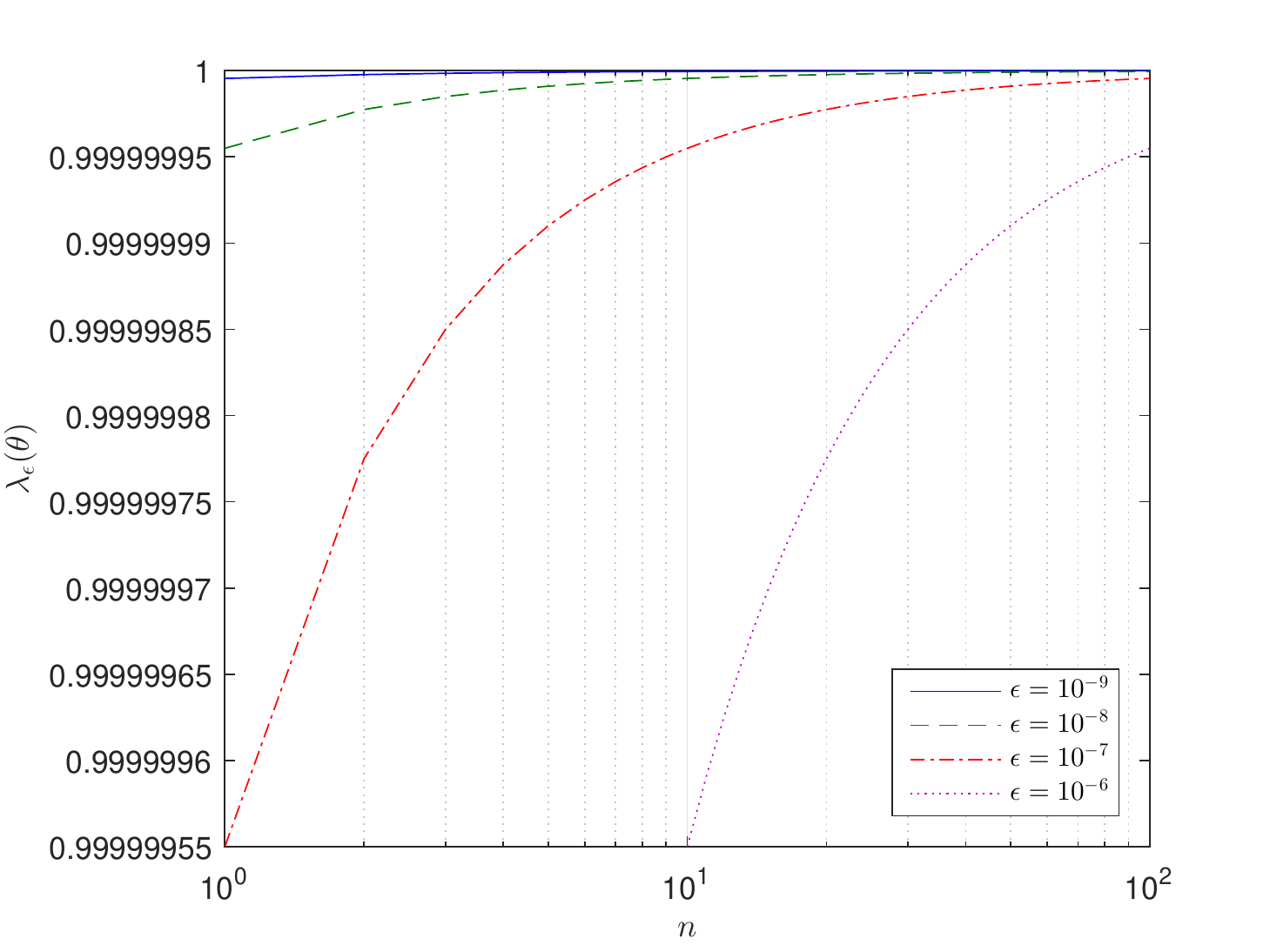}\label{results1a}}
\hfil
\subfloat[Probably correct reliability constraint ]{\includegraphics[scale=0.61]{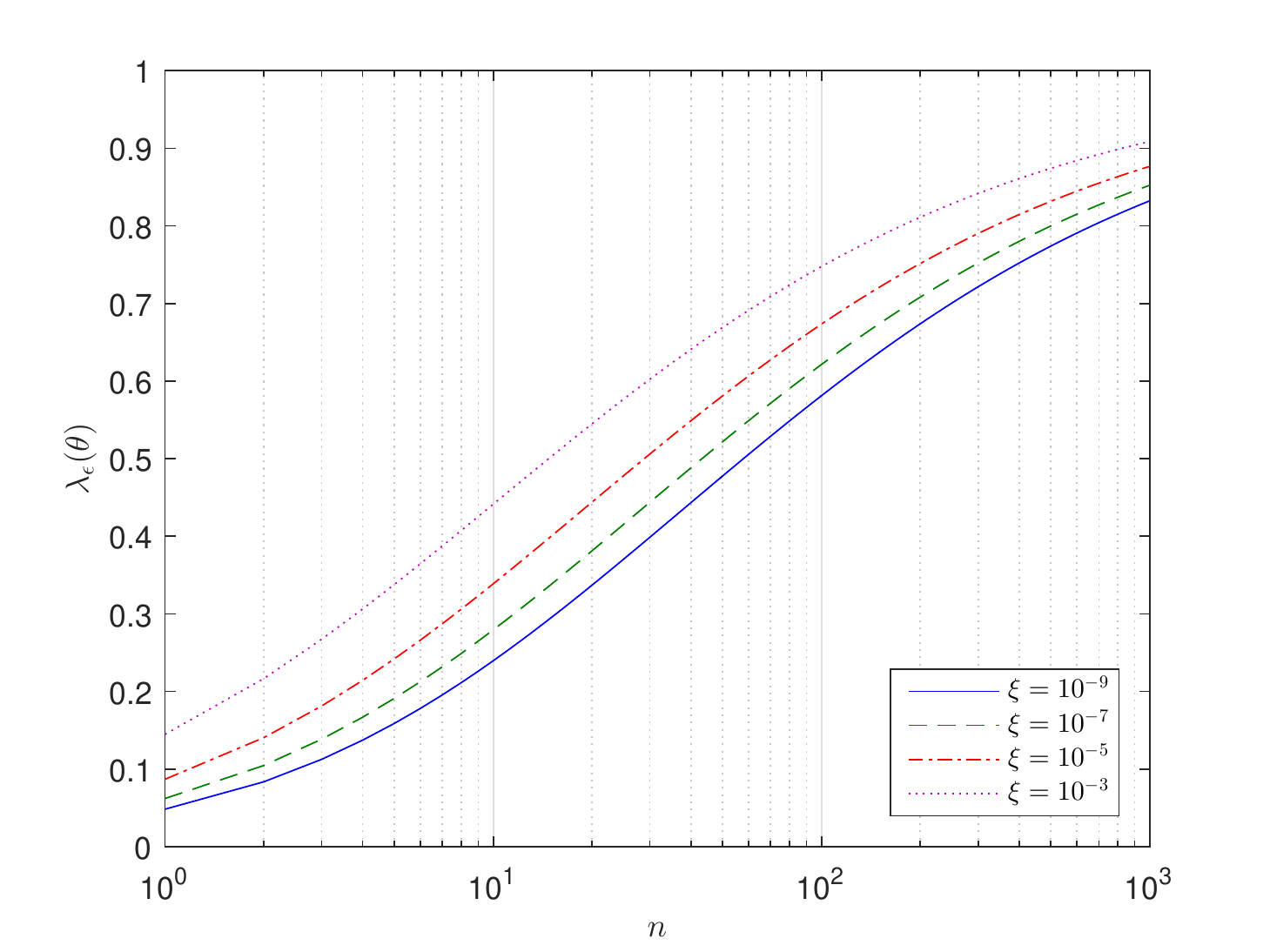}\label{results1b}}
\caption{Parametric rate-selection under Rayleigh channel fading with average power $\theta=10$.}
\label{stat_urllc_mk}
\end{figure*}

\subsubsection{Probably Correct Reliability (PCR)}

We consider another, more restrictive reliability constraint that effectively controls the higher order moments of the outage probability via the concept of \emph{meta-probability} \cite{Haenggi2016}; namely, we require:
\begin{equation}
\sup_{\theta} \text{Pr}[\text{Pr}[R(X^n)>\text{log}_2(1+Y)|X^n]>\epsilon]\leq\xi.
\end{equation}
In other words, we limit the probability that the outage probability given $X^n$ is larger than $\epsilon$.
Intuitively, $\xi$ is an upper limit on the willingness of the system to tolerate outages larger than $\epsilon$.
So, the probability of outage is guaranteed to be equal or less than $\epsilon$ with probability larger than $\xi$; again note that this does not guarantee that for specific realization of $X^n$ the outage probability will not be larger than $\epsilon$. 
This approach is more suitable for static environments where channel training is done infrequently.
In such circumstances, the average channel power estimate is used to set the rate for multiple, possibly many future transmission cycles; obviously, the transmitter here needs to be more conservative. 

Using the Rayleigh-channel assumption and the rate-selection function \eqref{eq:rate_selection}, we obtain the following simple result:
\begin{align}\label{eq:const2}
& \sup_{\theta} \text{Pr}[\text{Pr}[R(X^n)>\text{log}_2(1+Y)|X^n]>\epsilon] = \\
& = 1 - \frac{\gamma\left(n,n\frac{\log(1-\epsilon)}{\log(1-\varepsilon_n)}\right)}{(n-1)!},
\end{align}
with $\gamma(\cdot,\cdot)$ denoting the lower incomplete gamma function.
By choosing $\varepsilon_n$ satisfying
\begin{align}\label{eq:varen_constII_rayleigh}
1 - \frac{\gamma\left(n,n\frac{\log(1-\epsilon)}{\log(1-\varepsilon_n)}\right)}{(n-1)!} \leq \xi,
\end{align}
we obtain a rate-selection function that satisfies \eqref{eq:const2} for any $\theta$.
Note that in the specific study of Rayleigh channel, the choice of rate using PCR approach does not depend on $\theta$.
This is a convenient result, stating that in case of Rayleigh channel, the transmitter only needs to ensure that the rate does not violate the maximum allowed tolerance $\xi$; such rate will be valid for any $\epsilon$. 


\subsubsection{Evaluation}

We evaluate both approaches w.r.t. the ratio between the average achievable throughput using rate $R(X^n)$ and the optimal throughput given that the distribution is known 
\begin{equation}
\lambda_{\epsilon}(\theta) = \frac{\mathbb{E}\mathopen{}\left[R(X^n)1_{R(X^n)\leq\log_2(1+Y)}\right]}{R_{\epsilon}(\theta)(1-\epsilon)}.
\end{equation}
We evaluate the average throughput via Monte-Carlo simulation with $K$ trials by simply averaging the rates $R(x^n)$ provided that $R(x^n)\leq\log_2(1+y)$ for any pair of realizations $y,x^n$.
We depict $\lambda_{\epsilon}(\theta)$ as a function of $n$ for different $\epsilon$ and for $\theta = 10$ in Fig.~\ref{stat_urllc_mk}.
We see that $\lambda<1$ in both approaches, i.e. limited channel knowledge reduces the transmission rate.
We also observe that the rate-selection function is consistent, that is as $n$ grows large $\lambda\rightarrow 1$.
This reduction is dramatically visible for the PCR constraint;
this is also expected as \eqref{eq:const2} is significantly more restrictive than \eqref{eq:const1}.
Hence, using meta-probability, although providing stricter and more firm reliability guarantees, reduces the average rate significantly.
Beside, the rate converges significantly slower to the respective $\epsilon$-outage capacity.

\subsection{Alternative channel models}

We note that the rate-selection function cannot be always guaranteed to be consistent.
Specifically, when relaying on parametric channel models, the rate is consistent only in the case when the true channel distribution adheres to the adopted model as above, where we assumed that the true channel envelope is Rayleigh distributed.
However, if the true channel differs even slightly from the assumed model (e.g. a small specular component is also present), the rate is not longer consistent and both the AR and PCR constraints will be violated, see \cite{angjel2018urlcstat} for in-depth discussions. This can be viewed as a general pitfall of parametric channel models; while they provide fast convergence, they are prone to significant bias which lead to inconsistent rate and severe reliability violations.
As an alternative, one can consider non-parametric channel modeling approaches. 
They indeed are guaranteed to give consistent rates under very mild channel restrictions but they require extensive training; in fact, the number of channel samples necessary to obtain non-negative transmission rate grows as \cite{angjel2018urlcstat}
\begin{equation}
    n\sim\frac{1}{\epsilon}.
\end{equation}
Finally, \cite{angjel2018urlcstat} suggests to use power law approximations of the channel tail as third alternative as they provide ``the best of the two worlds''; even though they do not guarantee rate consistency due to approximation error, they are significantly less biased than poor parametric models and they require reasonable training samples lengths.


\section{Related Work}
\label{sec:RelatedWork}

The body of literature on URLLC has been rapidly growing during the recent years.
Here we make a brief account of the literature related to the topics assessed in the paper, noting that this account is by no means exhaustive.


From the standardization perspective, the requirements of the URLLC ``usage scenario'' are originally defined in 3GPP document~\cite{TR38.913}.
Automotive URLLC cases, i.e. their service architectures and requirements, are elaborated in~\cite{TR22.886}.
Similarly, specification~\cite{TS22.261} elaborates service architectures and requirements for URLLC use cases belonging to Industry 4.0 and intelligent transportation systems (ITS).
Finally, a recent specification concerning 5G New Radio (NR) and Radio Access Network description~\cite{3GPPTS-38300} proposes some methods for the facilitation of URLLC services, elaborated later in this section.  


A detailed description of URLLC use cases that are expected to play important roles in future 5G networks can be found in~\cite{D2.1}; these include automotive, Industry 4.0, ITS and ad-hoc disaster and emergency relief.
Another take on emerging mission-critical services in cellular networks, such as tele-surgery, ITS and industrial automation, is given in~\cite{CASHBLV2018}.
Live audio production, another potential URLLC use case, is described in~\cite{PHSS2018}; notably, this use case emphasizes isochronous communication that is not supported by 4G and the article elaborates the ways to achieve it in 5G networks.
A common feature of all mentioned all URLLC use cases is that they are described by a single combination of target reliability and latency parameters.
In contrast, \cite{popovski2014ultra} proposes a more general URLLC service model with reliable-service composition, by which target reliability performance progressively increases with latency.


Another line of works deals with general treatment of wireless URLLC communications.
A survey of challenges and methods related to support of low-latency wireless communications is presented in~\cite{JGFMPZF2018}.
An assessment of the sources of diversity and their exploitation to enable wireless URLLC networking is given in~\cite{PNSCSTBKKS2018}. 
The challenges of high-performance wireless communications for industrial control are in focus of~\cite{LPD2017}.
The study the fundamental energy-latency tradeoff in URLLC systems employing incremental-redundancy hybrid automatic repeat request in a specific context of point-to-point wireless connectivity is presented in~\cite{AKC2018}.


Traditionally, channel models have been used to study average or cell edge channel conditions. In the case of URLLC, where reliability requirements are in the order of $10^{-5} - 10^{-9}$, it is the extreme tail of the channel model distribution that is important. In~\cite{eggers2017wireless}, the URLLC level behavior of common wireless channel models is investigated. The authors find that in many cases, a simple power law model with fitted parameters can sufficiently characterize the tail. In practical systems, where only limited channel knowledge is available and thus the specific channel model is unknown, the model uncertainty will inevitably impact the overlying communication protocol. In \cite{swamy2018wireless}, the most critical dimensions of uncertainty are identified and their impact analyzed for two examples of cooperative communication protocols. A different perspective is taken in \cite{bennis2018ultra}, where, first, relevant metrics and key enables of URLLC are discussed, where after mathematical tools from different scientific disciplines are proposed for evaluating the URLLC properties of different wireless communication system applications. While traditionally dependability metrics such as availability and reliability are expressed as functions of time, \cite{mendis2017achieving} proposes an evaluation framework for extending such analyses in the space domain, specifically considering ultra-reliable heterogeneous and homogeneous cellular communication systems.


A vast number of papers discuss cellular access networking for URLLC services, we mention only a handful of them that explicitly deal with uplink communications.
Some guidelines on design and optimization of LTE and 5G new-radio interfaces related to numerology, robust transmission modes and connection management for URLLC services are presented in~\cite{SWDBK2018}; the overall conclusion is that the spectral efficiency is reduced with respect to services without latency and/or reliability constraints.
Along this lines, 3GPP document~\cite{3GPPTS-38300} specifies the use of appropriate numerology and transmission durations for uplink resource reservations for URLLC traffic, such that adequate subcarrier spacing and/or short-duration transmission can be achieved. 
An overview of high-level optimization of radio-resource management for both uplink and downlink using the tools of network calculus and effective bandwidth is made in~\cite{SYQ2017}.
Finding a scheduling policy when that allows multiple users to meet a target reliable-latency objective is the topic of~\cite{DPAK2018}, where the dynamic programming and knapsack-inspired approaches are used to find optimal and computationally-efficient suboptimal policies, respectively.

A number of works address advanced grant-free and grant-based access schemes for cellular access networks.
The benefits of multi-user detection in radio-access uplink are presented in~\cite{SC2017}.
A scheme that exploit multi-user detection in combination with grant-free access with proactive sending of packet replicas is proposed in~\cite{STLU2018}.
A grant-free scheme for batch arrivals that exploits replicas and SIC-based receiver and offers latency-reliability guarantees is introduced in~\cite{SLP2017}. 
Coordinated sharing of uplink resources in a grant-free manner, where the arrived users transmit packet replicas according to  predefined activity patterns is proposed in~\cite{KMBP2018}; the performance of the scheme is analyzed for both MMSE- and SIC-based receivers.
Comparison of grant-free schemes with proactive and reactive sending of replicas is made in~\cite{JABPMKM2017}; under the assumptions of an ideal control channel, perfect channel estimation, MMSE-interference rejection combining receiver, and target reliability of $1-10^{-5}$, it is shown that latency of grant-free schemes is generally lower than of the grant-based ones, while the choice between proactive and reactive grant-free schemes should be based on the load of the access network.
In regards to the grant-based, four-step access, the work~\cite{SIJPLU2018} proposes a control channel design that increases the reliability of signaling exchanges.
Finally, for the sake of completeness, we note that the 3GPP specification~\cite{3GPPTS-38300} in essence inherits the uplink access procedures from LTE, see remarks about LTE connection establishment in Section~\ref{sec:four-step}.
Successfully connected device can be granted periodically occurring resources, which may be an option for devices with predictable activity; otherwise the reserved resources will not be used efficiently.
It should also be noted that 3GPP at this point does not specify grant-free schemes for mobile cellular access.


Multiple antenna systems appear as a natural enabler for URLLC as multiple antenna communications provide high SNR and diversity links as well as spatial multiplexing capability. Those properties contribute to increasing the reliability or the latency or both. In spite of their obvious advantages, however, those mechanisms are still relatively unexplored.
A statistical characterization of multiple-input, single-output systems under statistical delay constraints is investigated in~\cite{AK2018}, using the tools of stochastic network calculus.
In \cite{feasibility_large_arrays_urllc}, for the uplink of a single user massive MIMO system with 64 antennas at the base station, diverse multi-antenna schemes are tested: coherent and non-coherent transceivers, transceivers assuming that the channel is unknown at the transmitter and using space-time codes, where the preference is given to non-coherent transceivers. In {\cite{massive_MIMO_tactile_internet}}, the accent is put on the processing delay caused by multiple antenna processing at a base station in a multi-user massive MIMO system. In \cite{urllc_mmwave_massivemimo}, a multi-user massive MIMO system network is optimized under a probabilistic constraint on the queue size to satisfy URLLC requirements. 

One central question in multi-antenna system is the acquisition of instantaneous CSI. It is one of the most severe limitations to achieve URLLC when exploiting multiple antennas in a mobile environment constrained by channel coherence time as well as extreme latency requirements. The most critical acquisition occurs at the antenna array when the CSI is used for transmission mode (CSIT). In frequency division duplex (FDD) systems, CSIT acquisition requires a feedback loop from the terminals inducing a significant latency as the number of links to report is large. In a massive MIMO system below 6~GHz where the terminals have a small number of antennas, this concerns the BS in downlink transmission.  In time division duplex (TDD) systems, latency can still be reduced by exploiting channel reciprocity {\cite{massivemimo_perf_imperfect_channel}}, but remains critical. In a mmWave system, with potentially many antennas at the terminal, the issue concerns both side of the communication links.

Acquisition of the CSI at the receiver is usually perceived as less critical than at the transmitter as the delay between channel estimation and data detection is short. However, extreme cases of mobility at the user side might require an alternative to coherent detection, especially if URLLC is the target. Hence, non-coherent detection methods can be an asset in that case. A particularly simple method~\cite{noncoherent_design_performance} that greatly benefits from the presence of a massive number of antennas is based on energy detection at the uplink of a massive MIMO system. The principle is to send a single stream of data, collect and aggregate the energy from all antennas. Detection is performed based on the average channel energy across the antenna array, which tends to a deterministic quantity for localized movements of the user and is therefore much more robust to user mobility than coherent detection.  In addition, an efficient constellation design has been proposed in~\cite{dual_stage_non_coh} that is able to benefit from the advantages of coherent communications at low mobility while switching to energy detection to ensure reliable communications at high mobility.

It is well-known from reliability engineering that ultra-high reliability can be effectively achieved through the parallel use of independent system components. In communication systems this translates to using multiple channels in parallel to achieve redundancy. 3GPP NR rel. 15 specifies use of Packet Duplication for URLLC services, where two independent transmission paths, from two different BSs, are used simultaneously to increase reliability and lower latency. Due to the involvement of multiple BSs, packet duplication requires modifications in the network architecture \cite{rao2018packet,aijaz2018packet}. Packet Duplication can be achieved both through Dual Connectivity and Carrier Aggregation, which are compared in \cite{rao2018packet}. In \cite{wolf2017diversity}, an information theoretic study of the achievable gain of using joint decoding in a multi-connectivity setting over traditional MSC and MRC schemes is presented. For these schemes, the diversity multiplexing tradeoff that allows to trade outage-off outage probability and system throughput, is investigated. While the latency is not explicitly quantified in this work, the use of short packets and demonstrated reduction of outage probability enables URLLC. Finally, field trial measurements are used to demonstrate the practical performance.
An extension of multi-connectivity is to complement BS links with D2D links, whereby UEs function as relays. A mathematical evaluation framework based on short block length regime for such D2D extended systems is proposed in \cite{she2018improving} and where DF and AF relay strategies are compared to traditional BS-oriented MC.
While multi-connectivity is typically assuming LTE or 5G links in a 3GPP system, the authors in \cite{chandrashekar20165g} propose network architecture for enabling multi-RAT multi-connectivity, thereby allowing for example Wi-Fi or LTE-LAA to be exploited for multi-connectivity. As shown in \cite{nielsen2017ultra}, a combination of several different types of communication links can be flexibly used so as to optimize for a service-specific latency/reliability/bandwidth consumption tradeoff. 


\section{Discussion and Outlook}
\label{sec:conclusion} 

In this paper we have discussed the principles of wireless access for Ultra-Reliable Low Latency Communication (URLLC). We have used a communication-theoretic framework to provide discussion on the fundamental tradeoffs. This was followed by elaboration on the important elements in access protocols. Two specific technologies were considered in the context of ultra-reliable communication, massive MIMO and multi-connectivity (interface diversity). We have also touched upon the important question about the proper statistical methodology for designing and assessing ultra-reliable communication. 

In the final remarks, we turn towards the issue of coupling high reliability with low latency, as it is done in the context of 5G. Relaxing the latency requirements towards a long term, e.~g. beyond $10$ or $50$ ms, opens the design space for solutions that have good system level characteristics, such as coexistence with the other 5G services, or exhibit a higher energy efficiency. It should also be noted that the latency requirements on the wireless link can be reduced by adopting a holistic system design. For example, requiring 1 ms from the wireless link, while allowing source compression procedures that introduce much larger delay, is certainly not the optimal approach. Hence, one is tempted to define new research problems related to joint source-channel coding and protocol design that are suited to meet end-to-end latency and reliability requirements.

\section*{Acknowledgment}
The work has partly been supported by the European Research Council (ERC Consolidator Grant nr. 648382 WILLOW), and partly by the Horizon 2020 project ONE5G (ICT-760809).


\bibliographystyle{IEEEtran}
\bibliography{./bibliography}

\begin{thebibliography}{10}
\providecommand{\url}[1]{#1}
\csname url@samestyle\endcsname
\providecommand{\newblock}{\relax}
\providecommand{\bibinfo}[2]{#2}
\providecommand{\BIBentrySTDinterwordspacing}{\spaceskip=0pt\relax}
\providecommand{\BIBentryALTinterwordstretchfactor}{4}
\providecommand{\BIBentryALTinterwordspacing}{\spaceskip=\fontdimen2\font plus
\BIBentryALTinterwordstretchfactor\fontdimen3\font minus
  \fontdimen4\font\relax}
\providecommand{\BIBforeignlanguage}[2]{{%
\expandafter\ifx\csname l@#1\endcsname\relax
\typeout{** WARNING: IEEEtran.bst: No hyphenation pattern has been}%
\typeout{** loaded for the language `#1'. Using the pattern for}%
\typeout{** the default language instead.}%
\else
\language=\csname l@#1\endcsname
\fi
#2}}
\providecommand{\BIBdecl}{\relax}
\BIBdecl

\bibitem{popovski2014ultra}
P.~Popovski, ``Ultra-reliable communication in {5G} wireless systems,'' in
  \emph{1st Int. Conf. on 5G for Ubiquitous Connectivity (5GU)}.\hskip 1em plus
  0.5em minus 0.4em\relax IEEE, 2014, pp. 146--151.

\bibitem{PNSCSTBKKS2018}
P.~Popovski, J.~J. Nielsen, C.~Stefanovic, E.~d.~Carvalho, E.~Strom, K.~F.
  Trillingsgaard, A.~Bana, D.~M. Kim, R.~Kotaba, J.~Park, and R.~B. Sorensen,
  ``{Wireless Access for Ultra-Reliable Low-Latency Communication: Principles
  and Building Blocks},'' \emph{IEEE Network}, vol.~32, no.~2, pp. 16--23, Mar.
  2018.

\bibitem{durisi2015towards}
G.~Durisi, T.~Koch, and P.~Popovski, ``Towards massive, ultra-reliable, and
  low-latency wireless communication with short packets,'' \emph{Proc. IEEE},
  vol. 104, no.~9, pp. 1711 -- 1726, sep 2016.

\bibitem{porter2014smart}
M.~E. Porter and J.~E. Heppelmann, ``How smart, connected products are
  transforming competition,'' \emph{Harvard Business Review}, vol.~92, no.~11,
  pp. 64--88, 2014.

\bibitem{TR38.913}
{3GPP}, ``3gpp tr 38.913 v15.0.0: Study on scenarios and requirements for next
  generation access technologies; (release 15),'' Tech. Rep., Jun. 2018.

\bibitem{D2.1}
\BIBentryALTinterwordspacing
{E2E-aware Optimizations and advancements for Network Edge of 5G New Radio
  (ONE5G)}, ``{Deliverable D2.1: Scenarios, KPIs, use cases and baseline system
  evaluation},'' Tech. Rep., Nov. 2017, {Accessed 25-August-2018}. [Online].
  Available:
  \url{https://one5g.eu/wp-content/uploads/2017/12/ONE5G_D2.1_finalversion.pdf}
\BIBentrySTDinterwordspacing

\bibitem{TR22.886}
{3GPP}, ``3gpp tr 22.886 v16.0.0: Study on enhancement of 3gpp support for 5g
  v2x services (release 16),'' Tech. Rep., Jun. 2018.

\bibitem{TS22.261}
------, ``3gpp ts 22.261 v16.4.0: Service requirements for the 5g system; stage
  1 (release 16),'' Tech. Rep., Jun. 2018.

\bibitem{ericsson}
\BIBentryALTinterwordspacing
Ericsson, ``{Manufacturing reengineered: robots, 5G and the Industrial IoT},''
  \emph{Ericsson Business Review}, vol.~4, 2015, {Accessed 02-October-2018}.
  [Online]. Available:
  \url{https://www.ericsson.com/assets/local/publications/ericsson-business-review/issue-4--2015/ebr-issue4-2015-industrial-iot.pdf}
\BIBentrySTDinterwordspacing

\bibitem{siemens}
\BIBentryALTinterwordspacing
Siemens, ``{5G communication networks: Vertical industry requirements},'' Tech.
  Rep., 2016, {Accessed 02-October-2018}. [Online]. Available:
  \url{http://www.virtuwind.eu/_docs/Siemens_PositionPaper_5G_2016.pdf}
\BIBentrySTDinterwordspacing

\bibitem{CASHBLV2018}
H.~Chen, R.~Abbas, P.~Cheng, M.~Shirvanimoghaddam, W.~Hardjawana, W.~Bao,
  Y.~Li, and B.~Vucetic, ``{Ultra-Reliable Low Latency Cellular Networks: Use
  Cases, Challenges and Approaches},'' \emph{IEEE Commun. Mag.}, Sep. 2018.

\bibitem{HWWTAHAA2016}
B.~Holfeld, D.~Wieruch, T.~Wirth, L.~Thiele, S.~A. Ashraf, J.~Huschke,
  I.~Aktas, and J.~Ansari, ``{Wireless Communication for Factory Automation: an
  opportunity for LTE and 5G systems},'' \emph{IEEE Commun. Mag.}, vol.~54,
  no.~6, pp. 36--43, Jun. 2016.

\bibitem{M.2134}
\BIBentryALTinterwordspacing
ITU-R, ``{REPORT ITU-R M.2134: Requirements related to technical performance
  for IMT-Advanced radio interface(s)},'' Tech. Rep., 2008, {Accessed
  06-October-2018}. [Online]. Available:
  \url{https://www.itu.int/pub/R-REP-M.2134}
\BIBentrySTDinterwordspacing

\bibitem{bennis2018ultra}
M.~Bennis, M.~Debbah, and H.~V. Poor, ``Ultra-reliable and low-latency wireless
  communication: Tail, risk and scale,'' \emph{Proc. IEEE}, vol. 106, no.~10,
  pp. 1834 -- 1853, Oct. 2018.

\bibitem{ETSI2014LTN}
\BIBentryALTinterwordspacing
ETSI, ``{Low Throughput Networks (LTN); Functional Architecture},'' in
  \emph{ETSI GS LTN 002 V1.1.1}, 2014. [Online]. Available:
  \url{http://www.etsi.org/deliver/etsi_gs/LTN/001_099/002/01.01.01_60/
  gs_LTN002v010101p.pdf}
\BIBentrySTDinterwordspacing

\bibitem{finite_blocklength_polyanskiy}
Y.~Polyanskiy, H.~V. Poor, and S.~Verdú, ``Channel coding rate in the finite
  blocklength regime,'' \emph{IEEE Trans. Inf. Theory}, vol.~56, no.~5, Apr.
  2010.

\bibitem{bana_short_pkt_tradeoff}
A.~Bana, K.~F. Trillingsgaard, P.~Popovski, and E.~de~Carvalho, ``Short packet
  structure for ultra-reliable machine-type communication: Tradeoff between
  detection and decoding,'' in \emph{2018 IEEE Int. Conf. on Acoustics, Speech
  and Signal Processing (ICASSP)}, Apr. 2018, pp. 6608--6612.

\bibitem{trillingsgaard2017downlink}
K.~F. Trillingsgaard and P.~Popovski, ``Downlink transmission of short packets:
  Framing and control information revisited,'' \emph{IEEE Trans. Commun.},
  vol.~65, no.~5, pp. 2048--2061, 2017.

\bibitem{tuninetti2018scheduling}
D.~Tuninetti, B.~Smida, N.~Devroye, and H.~Seferoglu, ``Scheduling on the
  gaussian broadcast channel with hard deadlines,'' in \emph{2018 IEEE Int.
  Conf. on Commun. (ICC)}.\hskip 1em plus 0.5em minus 0.4em\relax IEEE, 2018,
  pp. 1--7.

\bibitem{ostman2017short}
J.~{\"O}stman, G.~Durisi, E.~G. Str{\"o}m, M.~C. Co{\c{s}}kun, and G.~Liva,
  ``Short packets over block-memoryless fading channels: Pilot-assisted or
  noncoherent transmission?'' \emph{IEEE Trans. Commun.}, Oct. 2018.

\bibitem{ferrante2018pilot}
G.~C. Ferrante, J.~Ostman, G.~Durisi, and K.~Kittichokechai, ``Pilot-assisted
  short-packet transmission over multiantenna fading channels: A 5g case
  study,'' in \emph{2018 52nd Annu. Conf. on Inf. Sci. and Syst. (CISS)}.\hskip
  1em plus 0.5em minus 0.4em\relax IEEE, 2018, pp. 1--6.

\bibitem{berardinelli2015design}
G.~Berardinelli, K.~Pedersen, F.~Frederiksen, and P.~Mogensen, ``On the design
  of a radio numerology for 5g wide area,'' in \emph{Proc. 11th Int. Conf.
  Wireless Mobile Commun.(ICWMC)}, 2015, pp. 13--18.

\bibitem{ITU-T}
\BIBentryALTinterwordspacing
``{ITU-T Recommendation G.8260},'' {Accessed 17-August-2018}. [Online].
  Available: \url{https://www.itu.int/rec/T-REC-G.8260-201508-I/}
\BIBentrySTDinterwordspacing

\bibitem{PSLP2014}
E.~Paolini, C.~Stefanovic, G.~Liva, and P.~Popovski, ``Coded random access:
  {H}ow coding theory helps to build random access protocols,'' \emph{IEEE
  Commun. Mag.}, vol.~53, no.~6, pp. 144--150, Jun. 2015.

\bibitem{GSP2018}
J.~Goseling, C.~Stefanovic, and P.~Popovski, ``{Sign-Compute-Resolve for Tree
  Splitting Random Access},'' \emph{IEEE Trans. Info. Theory}, vol.~64, no.~7,
  pp. 5261--5276, Jul. 2018.

\bibitem{LLYPSC2018}
L.~Liu, E.~G. Larsson, W.~Yu, P.~Popovski, C.~Stefanovic, and E.~D. Carvalho,
  ``Sparse signal processing for grant-free massive connectivity: A future
  paradigm for random access protocols in the internet of things,'' \emph{IEEE
  Signal Process. Mag.}, vol.~35, no.~5, pp. 88 -- 99, Sep. 2018.

\bibitem{TS36.321}
{3GPP}, ``3gpp ts 36.321 v12.5.0: Evolved universal terrestrial radio access
  (e-utra); medium access control (mac) protocol specification; (release 15),''
  Tech. Rep., Apr. 2015.

\bibitem{madueno2016assessment}
G.~C. Madue{\~n}o, J.~J. Nielsen, D.~M. Kim, N.~K. Pratas,
  {\v{C}}.~Stefanovi{\'c}, and P.~Popovski, ``Assessment of lte wireless access
  for monitoring of energy distribution in the smart grid,'' \emph{IEEE J. Sel.
  Areas Commun}, vol.~34, no.~3, pp. 675--688, 2016.

\bibitem{M1972}
J.~Massey, ``{Optimum Frame Synchronization},'' \emph{IEEE Trans. Commun.},
  vol.~20, no.~2, pp. 115--119, Apr. 1972.

\bibitem{R1994}
P.~Robertson, ``{Optimum frame synchronization of preamble-less packets
  surrounded by noise with coherent and differentially coherent
  demodulation},'' in \emph{Proceedings of ICC/SUPERCOMM'94 - 1994
  International Conference on Communications}, May 1994, pp. 874--879 vol.2.

\bibitem{SB2012}
C.~Stefanovic and D.~Bajic, ``{On the Search for a Sequence from a Predefined
  Set of Sequences in Random and Framed Data Streams},'' \emph{IEEE Trans.
  Commun.}, vol.~60, no.~1, pp. 189--197, Jan. 2012.

\bibitem{SJ1988}
M.~N. Al-Subbagh and E.~V. Jones, ``Optimum patterns for frame alignment,''
  \emph{IEE Proceedings F - Communications, Radar and Signal Processing}, vol.
  135, no.~6, pp. 594--604, Dec. 1988.

\bibitem{GG2009}
S.~W. Golomb and G.~Gong, \emph{Signal Design for Good Correlation}.\hskip 1em
  plus 0.5em minus 0.4em\relax Cambridge University Press, 2009.

\bibitem{APMS2017}
A.~Azari, P.~Popovski, G.~Miao, and C.~Stefanovic, ``{Grant-Free Radio Access
  for Short-Packet Communications over 5G Networks},'' in \emph{IEEE GLOBECOM
  2017}, Dec. 2017.

\bibitem{five_disruptive_tech_5g}
F.~Boccardi, R.~W. Heath, A.~Lozano, T.~L. Marzetta, and P.~Popovski, ``{Five
  disruptive technology directions for 5G},'' \emph{{IEEE} Commun. Mag.},
  vol.~52, no.~2, 2014.

\bibitem{massivemimo_massive_connect}
L.~Liu and W.~Yu, ``{Massive device connectivity with massive MIMO},''
  \emph{Int. Symp. Inf. Theory (ISIT)}, Jun. 2017.

\bibitem{feasibility_large_arrays_urllc}
S.~R. Panigrahi, N.~Bj{\"o}rsell, and M.~Bengtsson, ``{Feasibility of Large
  Antenna Arrays towards Low Latency Ultra Reliable Communication},''
  \emph{IEEE Int. Conf. on Industrial Technology (ICIT)}, Mar. 2017.

\bibitem{urllc_mmwave_massivemimo}
T.~K. Vu, C.-F. Liu, M.~Bennis, M.~Debbah, M.~Latva-aho, and C.~S. Hong,
  ``{Ultra-Reliable and Low Latency Communication in mmWave-Enabled Massive
  MIMO Networks},'' \emph{{IEEE} Commun. Lett.}, no.~99, May 2017.

\bibitem{favorable_propag}
H.~Q. Ngo, E.~G. Larsson, and T.~L. Marzetta, ``{Aspects of favorable
  propagation in Massive MIMO},'' \emph{Proc. $22^\text{nd}$ European Sign.
  Proc. Conf. (EUSIPCO)}, 2014.

\bibitem{massive_MIMO_tactile_internet}
W.~Tarneberg, M.~Karaca, A.~Robertsson, F.~Tufvesson, and M.~Kihl, ``{Utilizing
  Massive MIMO for the Tactile Internet: Advantages and Trade-Offs},''
  \emph{{IEEE} Int. Conf. Sensing, Comm. and Networking (SECON Workshops)},
  2017.

\bibitem{Boccardi09}
F.~Boccardi, H.~Huang, and M.~Trivellato, ``A near-optimum precoding technique
  for downlink multi-user mimo transmissions,'' \emph{Bell Labs Technical
  Journal}, vol.~13, no.~4, pp. 79--95, Winter 2009.

\bibitem{JSDM13}
A.~Adhikary, J.~Nam, J.-Y. Ahn, and G.~Caire, ``Joint spatial division and
  multiplexing — the large-scale array regime,'' \emph{IEEE Trans. Inf.
  Theory}, vol.~59, no.~10, pp. 6441--6463, 2013.

\bibitem{Chowdhury17}
M.~Chowdhury, A.~Manolakos, and A.~Goldsmith, ``Multiplexing and diversity
  gains in noncoherent massive mimo systems,'' \emph{IEEE Trans. Wireless
  Commun.}, vol.~16, no.~1, pp. 265--277, Jan 2017.

\bibitem{SWDBK2018}
{J. Sachs and G. Wikstrom and T. Dudda and R. Baldemair and K. Kittichokechai},
  ``{5G Radio Network Design for Ultra-Reliable Low-Latency Communication},''
  \emph{IEEE Network}, vol.~32, no.~2, pp. 24--31, Mar. 2018.

\bibitem{nielsen2017ultra}
J.~J. Nielsen, R.~Liu, and P.~Popovski, ``{Ultra-reliable low latency
  communication (URLLC) using interface diversity},'' \emph{IEEE Trans.
  Commun.}, vol.~66, no.~3, pp. 1322 -- 1334, Mar. 2018.

\bibitem{Haenggi2016}
M.~Haenggi, ``The meta distribution of the {SIR} in poisson bipolar and
  cellular networks,'' \emph{IEEE Trans. Wireless Commun.}, vol.~15, no.~4, pp.
  2577--2589, April 2016.

\bibitem{angjel2018urlcstat}
M.~Angjelichinoski, K.~Fl{\o}e~Trillingsgaard, and P.~Popovski, ``A statistical
  learning approach to ultra-reliable low latency communication,'' \emph{arXiv
  preprint arXiv:1809.05515}, 2018.

\bibitem{3GPPTS-38300}
{3GPP}, ``Nr; overall description; stage-2,'' Tech. Rep., Jun 2018.

\bibitem{PHSS2018}
J.~Pilz, B.~Holfeld, A.~Schmidt, and K.~Septinus, ``{Professional Live Audio
  Production: A Highly Synchronized Use Case for 5G URLLC Systems},''
  \emph{IEEE Network}, vol.~32, no.~2, pp. 85--91, Mar. 2018.

\bibitem{JGFMPZF2018}
X.~Jiang, H.~S. Ghadikolaei, G.~Fodor, Z.~P. E.~Modiano, M.~Zorzi, and
  C.~Fischione, ``{Low-latency Networking: Where Latency Lurks and How to Tame
  It},'' \emph{Proc. IEEE}, Aug. 2018.

\bibitem{LPD2017}
M.~Luvisotto, Z.~Pang, and D.~Dzung, ``{Ultra High Performance Wireless Control
  for Critical Applications: Challenges and Directions},'' \emph{IEEE
  Transactions on Industrial Informatics}, vol.~13, no.~3, pp. 1448--1459, Jun.
  2017.

\bibitem{AKC2018}
\BIBentryALTinterwordspacing
A.~Avranas, M.~Kountouris, and P.~Ciblat, ``{Energy-Latency Tradeoff in
  Ultra-Reliable Low-Latency Communication with Retransmissions},'' \emph{IEEE
  J. Sel. Areas Commun.}, Oct. 2018. [Online]. Available:
  \url{http://arxiv.org/abs/1805.01332}
\BIBentrySTDinterwordspacing

\bibitem{eggers2017wireless}
P.~C. Eggers, M.~Angjelichinoski, and P.~Popovski, ``Wireless channel modeling
  perspectives for ultra-reliable low latency communications,'' \emph{arXiv
  preprint arXiv:1705.01725}, 2017.

\bibitem{swamy2018wireless}
V.~N. Swamy, P.~Rigge, G.~Ranade, B.~Nikolic, and A.~Sahai, ``Wireless channel
  dynamics and robustness for ultra-reliable low-latency communications,''
  \emph{arXiv preprint arXiv:1806.08777}, 2018.

\bibitem{mendis2017achieving}
H.~K. Mendis and F.~Y. Li, ``Achieving ultra reliable communication in 5g
  networks: A dependability perspective availability analysis in the space
  domain,'' \emph{IEEE Commun. Lett.}, vol.~21, no.~9, pp. 2057--2060, 2017.

\bibitem{SYQ2017}
C.~She, C.~Yang, and T.~Q.~S. Quek, ``Radio resource management for
  ultra-reliable and low-latency communications,'' \emph{IEEE Commun. Mag.},
  vol.~55, no.~6, pp. 72--78, Jun. 2017.

\bibitem{DPAK2018}
A.~Destounis, G.~S. Paschos, J.~Arnau, and M.~Kountouris, ``{Scheduling URLLC
  users with reliable latency guarantees},'' in \emph{16th Int. Symp. on
  Modeling and Optimization in Mobile, Ad Hoc, and Wireless Networks (WiOpt)},
  May 2018, pp. 1--8.

\bibitem{SC2017}
S.~Saur and M.~Centenaro, ``{Radio Access Protocols with Multi-User Detection
  for URLLC in 5G},'' in \emph{European Wireless 2017; 23rd European Wireless
  Conf.}, May 2017.

\bibitem{STLU2018}
B.~Singh, O.~Tirkkonen, Z.~Li, and M.~A. Uusitalo, ``Contention-based access
  for ultra-reliable low latency uplink transmissions,'' \emph{IEEE Wireless
  Commun. Lett.}, vol.~7, no.~2, pp. 182--185, Apr. 2018.

\bibitem{SLP2017}
C.~Stefanovic, F.~Lazaro, and P.~Popovski, ``{Frameless ALOHA with
  Reliability-Latency Guarantees},'' in \emph{GLOBECOM 2017 - 2017 IEEE Global
  Commun. Conf.}, Dec. 2017.

\bibitem{KMBP2018}
R.~Kotaba, C.~N. Manchón, T.~Balercia, and P.~Popovski, ``{Uplink
  Transmissions in URLLC Systems With Shared Diversity Resources},'' \emph{IEEE
  Wireless Commun. Lett.}, vol.~7, no.~4, pp. 590--593, Aug. 2018.

\bibitem{JABPMKM2017}
T.~Jacobsen, R.~Abreu, G.~Berardinelli, K.~Pedersen, P.~Mogensen, I.~Z. Kovacs,
  and T.~K. Madsen, ``{System Level Analysis of Uplink Grant-Free Transmission
  for URLLC},'' in \emph{2017 IEEE Globecom Workshops (GC Wkshps)}, Dec. 2017.

\bibitem{SIJPLU2018}
H.~Shariatmadari, S.~Iraji, R.~Jantti, P.~Popovski, Z.~Li, and M.~A. Uusitalo,
  ``{Fifth-Generation Control Channel Design: Achieving Ultrareliable
  Low-Latency Communications},'' \emph{IEEE Veh. Technol. Mag.}, vol.~13,
  no.~2, pp. 84--93, Jun. 2018.

\bibitem{AK2018}
J.~Arnau and M.~Kountouris, ``{Delay performance of MISO wireless
  communications},'' in \emph{2018 16th International Symposium on Modeling and
  Optimization in Mobile, Ad Hoc, and Wireless Networks (WiOpt)}, May 2018, pp.
  1--8.

\bibitem{massivemimo_perf_imperfect_channel}
D.~Mi, M.~Dianati, L.~Zhang, S.~Muhaidat, and R.~Tafazolli, ``{Massive MIMO
  Performance with Imperfect Channel Reciprocity and Channel Estimation
  Error},'' \emph{IEEE Trans. Comm.}, no.~99, Mar. 2017.

\bibitem{noncoherent_design_performance}
L.~Jing, E.~D. Carvalho, P.~Popovski, and Àlex Oliveras~Martínez, ``{Design
  and Performance Analysis of Noncoherent Detection Systems With Massive
  Receiver Arrays},'' \emph{IEEE Trans. Signal Process.}, vol.~64, no.~19, pp.
  5000 -- 5010, Oct. 2016.

\bibitem{dual_stage_non_coh}
A.-S. Bana, M.~Angjelichinoski, E.~de~Carvalho, and P.~Popovski, ``Massive
  {MIMO} for ultra-reliable communications with constellations for dual
  coherent-noncoherent detection,'' \emph{22nd International ITG Workshop on
  Smart Antennas (WSA)}, Mar. 2018.

\bibitem{rao2018packet}
J.~Rao and S.~Vrzic, ``Packet duplication for urllc in 5g: Architectural
  enhancements and performance analysis,'' \emph{IEEE Network}, vol.~32, no.~2,
  pp. 32--40, 2018.

\bibitem{aijaz2018packet}
A.~Aijaz, ``Packet duplication in dual connectivity enabled 5g wireless
  networks: Overview and challenges,'' \emph{arXiv preprint arXiv:1804.01058},
  2018.

\bibitem{wolf2017diversity}
A.~Wolf, P.~Schulz, D.~{\"O}hmann, M.~D{\"o}rpinghaus, and G.~Fettweis,
  ``Diversity-multiplexing tradeoff for multi-connectivity and the gain of
  joint decoding,'' \emph{arXiv preprint arXiv:1703.09992}, 2017.

\bibitem{she2018improving}
C.~She, Z.~Chen, C.~Yang, T.~Q. Quek, Y.~Li, and B.~Vucetic, ``Improving
  network availability of ultra-reliable and low-latency communications with
  multi-connectivity,'' \emph{IEEE Trans. Commun}, Jun. 2018.

\bibitem{chandrashekar20165g}
S.~Chandrashekar, A.~Maeder, C.~Sartori, T.~H{\"o}hne, B.~Vejlgaard, and
  D.~Chandramouli, ``5g multi-rat multi-connectivity architecture,'' in
  \emph{IEEE Int. Conf. on Commun. Workshops (ICC Workshops)}.\hskip 1em plus
  0.5em minus 0.4em\relax IEEE, 2016, pp. 180--186.

\end{thebibliography}

\end{document}